\documentclass[twocolumn]{aastex631}

\graphicspath{{./}{figures/}}
\usepackage{amsmath}
\usepackage{longtable}
\usepackage{xcolor}
\usepackage{hyperref}
\usepackage{chngcntr}
\usepackage{booktabs}
\usepackage{needspace}
\usepackage{threeparttablex} 
\shorttitle{The Properties of Neutron Star Involved GRB Host Galaxies}
\shortauthors{Jeong and Im}

\begin{document}

\title{Host Galaxy Properties of Gamma-ray Bursts Involving Neutron Star Binary Mergers \\ and Its Impact on Kilonovae Rates}
\correspondingauthor{Myungshin Im}
\author{Mankeun Jeong}
\author{Myungshin Im}
\email{myungshin.im@gmail.com}
\affiliation{SNU Astronomy Research Center, Astronomy Program, Department of Physics and Astronomy,\\ Seoul National University, Gwanak-gu, Seoul 151-742, Korea}

\begin{abstract}

In the upcoming gravitational wave (GW) observing runs, identifying host galaxies is crucial as it provides essential redshift information and enables the use of GW events as standard sirens. However, pinpointing host galaxies remains challenging due to the large localization uncertainties and the rapidly fading nature of their optical counterparts. Analyzing the host galaxies of short gamma-ray bursts (sGRBs) offers an alternative approach to deepen our understanding of the environments where binary neutron stars (BNS) primarily merge. This study compiles archival photometric data for the host galaxies of 76 sGRBs and 4 hybrid GRBs that are long GRBs accompanied by kilonova-like signals. We use this data to evaluate their physical properties through spectral energy distribution (SED) fitting. To assess the characteristics of the host galaxies, we utilized a volume-limited sample ($z<0.5$) from the COSMOS field as a control group. Contrary to expectations that the BNS merger rate is proportional to host stellar mass, the short and hybrid GRB population appears less massive than the mass-weighted distribution of the control sample. Instead, we propose a formulation for the expected BNS merger rate from a galaxy as $\log(n_{\mathrm{BNS}}/\mathrm{Gyr}) = 0.86 \times \log(M_*/M_{\odot}) + 0.44 \times \log(\mathrm{sSFR}/\mathrm{yr}) + 0.857$, which optimally explains the deviation between the stellar mass distributions of the GRB host galaxies and the control sample. These insights provide a strategic framework for targeted GW follow-ups and enhance our ability to identify potential host galaxies for future GW events.
\end{abstract}


\section{Introduction}\label{sec:intro}

During the first three gravitational wave (GW) observing runs, and now into the fourth, the Advanced Laser Interferometer Gravitational-Wave Observatory (aLIGO; \citealt{2015CQGra..32g4001L}), the Advanced Virgo (aVirgo; \citealt{2015CQGra..32b4001A}), and the Kamioka Gravitational Wave Detector (KAGRA; \citealt{2021PTEP.2021eA101A}) collaborations have successfully measured the masses and distances of compact binary objects \citep{2016PhRvL.116f1102A, 2017ApJ...848L..12A, 2021ApJ...913L...7A}. In particular, the first binary neutron star (BNS) merger event, GW170817, confirmed the simultaneous emission of both GW and electromagnetic (EM) signals, evidenced by an accompanying kilonova and a short gamma-ray burst (sGRB) (e.g., \citealt{2017ApJ...848L..13A, 2017Natur.551...71T, 2017ApJ...848L..19C, 2017Sci...358.1565E, 2017Natur.551...67P, 2017Natur.551...75S}). Such multi-messenger astronomy (MMA) events allow for the identification of the host galaxy \citep{2017ApJ...848L..22B, 2017ApJ...848L..31H, 2017ApJ...848L..28L, 2017ApJ...849L..16I, 2017ApJ...848L..30P, 2017ApJ...848L..23F}, from which we can measure the redshift and utilize the GW event as standard sirens \citep{2017Natur.551...85A, 2017Natur.551...80K, 2021ApJ...909..218A, 2021arXiv211106445P}. Therefore, in upcoming GW observing runs, the discovery of optical counterparts and host galaxies for GW events will be as crucial as ever.


Identifying the host galaxy and EM counterpart of GW can be challenging due to the substantial localization uncertainty and rapidly fading nature of kilonova (e.g., \citealt{2017Natur.551...64A, 2019ApJ...876..128K, 2020LRR....23....3A}). A notable example is GW190425, a probable BNS merger detected during the LIGO/Virgo O3 run. This event had an exceptionally large 90\% credible localization area, spanning 7461 square degrees \citep{2020ApJ...892L...3A}. Despite extensive efforts, no EM counterpart has been identified \citep{2019ApJ...885L..19C, 2020MNRAS.497.1181C, 2024ApJ...960..113P}.

In the ongoing GW O4 observing run, it has been anticipated that improvements in detector sensitivity would reduce the localization uncertainty of BNS merger events to tens of square degrees \citep{2020LRR....23....3A}. However, due to delays in integrating Advanced Virgo into the GW network, the expected enhancements in performance have not been realized. Consequently, the localization uncertainty for BNS merger events has escalated to thousands of square degrees, as predicted under less optimal operational scenarios \citep{2014ApJ...795..105S, 2015ApJ...804..114B}.


These challenges underscore the essential role of understanding host galaxy environments for prioritizing and planning effective GW follow-up observations \citep{2016ApJ...829L..15S, 2017Sci...358.1556C, 2018MNRAS.479.2374D, 2021arXiv211006184D}. GW alert brokers will streamline GW follow-up efforts by optimizing observation strategies, including tiling or targeting observations, to efficiently cover localized areas \citep{2022arXiv221200531S, 2023ApJ...947...59L}. In this context, integrating galaxy-oriented observations from ground-based telescopes into these strategies can enhance the ability to formulate effective approaches, underlining a thorough understanding of the host galaxy environment \citep{2022MNRAS.517L...5R, 2022AJ....163..209S}.


Simulation studies utilizing binary evolutionary models and galactic hydrodynamics have suggested that stellar mass should be a crucial predictor of GW event rates in local galaxies \citep{2018MNRAS.481.5324M, 2019MNRAS.487....2M, 2020MNRAS.491.3419A, 2022MNRAS.509.1557C}. This insight is based on the premise that, over the extended delay time during which a binary system merges, the host galaxy itself undergoes evolution from galaxy mergers. This concept has been incorporated into resources like the Extended Galaxy List for the Advanced Detector Era (GLADE+), which serves as a reference catalog for identifying host galaxies \citep{2021arXiv211006184D}.


The host galaxies of sGRBs are useful to validate predictions about the BNS merger environment, as they are one of the MMA signals generated by the collision of either BNS or neutron star-black hole (NSBH) binaries \citep{1999A.A...344..573R, 2003MNRAS.345.1077R, 2006ApJ...641L..93F, 2011LRR....14....6S}. Earlier analyses of sGRB host galaxies revealed a mixed population of early-type and star-forming galaxies, which were reported to predominantly follow an older stellar population compared to long GRB host galaxies \citep{2010ApJ...725.1202L, 2014ARA.A..52...43B}. 

Recent deep optical/NIR survey observations have revealed additional faint and high redshift host galaxies, prompting a reevaluation of the sGRB host population \citep{2022MNRAS.515.4890O, 2022ApJ...940...56F}. Following this, a comprehensive statistical analysis of 67 sGRB host galaxies by \citet{2022ApJ...940...57N} demonstrated that their mass, metallicity, and star formation rate (SFR) are consistent with the star-forming main sequence. Intriguingly, high redshift sGRBs indicate a steeper delay-time distribution (DTD) than previously anticipated by binary evolution models \citep{2022ApJ...940L..18Z}. This evidence challenges established theoretical perspectives on BNS mergers, identifying the host galaxy NGC 4993 of the BNS merger as an atypical example.


In response to the discrepancies between sGRB host statistics and simulation predictions, our study aims to quantitatively evaluate the global properties of local sGRB host galaxies. Additionally, to address uncertainties surrounding sGRB classifications, we will examine cases where kilonova-like signals were detected alongside GRBs with long prompt emissions or extended emission features. These hybrid GRBs are believed to originate from compact binary mergers involving neutron stars \citep{2022Natur.612..228T, 2024Natur.626..742Y}. We will analyze the stellar mass and SFR distributions of these host galaxies using spectral energy distribution (SED) fitting. This analysis will enable us to assess how these distributions deviate from those of typical galaxies in the local universe. The extent of these deviations may be indicative of the contribution of BNS mergers to the overall merger rate. The BNS merger rate derived from this analysis will help prioritize galaxies within probable volumes for future GW events and assist in assessing their redshift probability distributions.

The structure of this paper is as follows. In Sections~\ref{sec:sample} and \ref{sec:data}, the processes of collecting sGRB samples and photometry data are presented. Section~\ref{sec:sedmodelling} describes the SED fitting procedure and the control sample definition. And, Section~\ref{sec:result} presents the distributions of sGRB host galaxies properties and derivation of the BNS merger rate expectation. In Section~\ref{sec:discussion}, we will discuss the meaning of the BNS merger rate, potential sample biases and some limitations in our study. Finally, in Section~\ref{sec:conclusion}, a summary and conclusion are presented. Throughout the paper, magnitudes are in ABmag with the extinction correction of \citet{2011ApJ...737..103S}, and all the errors correspond to 1$\sigma$. The cosmological parameters are adopted as $H_{0}$ = 70 km s$^{-1}$ Mpc$^{-1}$ , $\Omega_{\Lambda}$ = 0.73, and $\Omega_{M}$ = 0.27.

\section{Sample} \label{sec:sample}

\subsection{sGRB Host Galaxy Sample}

To compile a comprehensive dataset of sGRB host galaxies, we combined target lists from various sources. The majority of the host galaxies are associated with sGRBs detected by the \emph{Swift} Burst Alert Telescope (BAT; \citealt{2005SSRv..120..143B}), in conjunction with the X-ray Telescope (XRT) and Ultra-violet Telescope (UVOT) \citep{2005SSRv..120...95R, 2005SSRv..120..165B}. This is primarily due to the fact that while the localization region for a GRB itself is rather uncertain for precisely identifying the host galaxy, the positional uncertainties are significantly reduced to subarcsecond or a few arcseconds when an optical or X-ray counterpart is discovered by \textit{Swift}.

Due to the absence of definitive redshift data, the identification of a host galaxy requires deep observations and careful investigation of field galaxies around the burst location \citep{2002AJ....123.1111B, 2010ApJ...722.1946B, 2014MNRAS.437.1495T}. Recent host population studies, like \citet{2022MNRAS.515.4890O} and \citet{2022ApJ...940...56F}, have conducted imaging observations reaching a deep limit (down to $r\gtrsim$ 26 ABmag). Such efforts have enhanced the completeness of the host galaxy dataset, setting a robust background for subsequent statistical analyses.

The Broadband Repository for Investigating Gamma-ray burst Host Traits (BRIGHT\footnote{\url{https://bright.ciera.northwestern.edu/}}) serves as an archive for the photometric data of most sGRB host galaxies identified between 2005 and 2021 \citep{2022ApJ...940...56F}. Among the host galaxies cataloged there, 69 are conclusively associated with sGRBs and possess sufficient photometric data to conduct SED analysis \citep{2022ApJ...940...57N}.

To enlarge the sGRB host galaxy sample, we examined recent Gamma-ray Coordinates Network (GCN) reports. Notably, there are dedicated web-based repositories, such as GRBweb\footnote{\url{https://sites.astro.caltech.edu/grbox/grbox.php}} and GRBOX\footnote{\url{https://user-web.icecube.wisc.edu}}. These sites archive observational logs and discussions about post-burst detection, primarily from instruments like the \emph{Swift} BAT and the \emph{Fermi} Gamma-ray Burst Monitor (GBM; \citealt{2009ApJ...702..791M}). By evaluating the burst duration and its association with potential host galaxies, we identified additional host candidates. 

\begin{figure}
\centering
\includegraphics[width=\columnwidth]{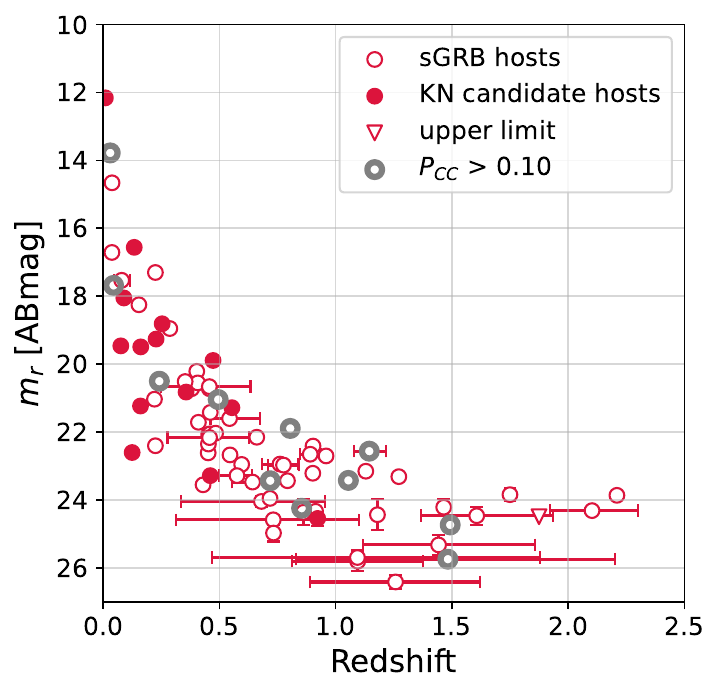}
\caption{The $r$-band apparent magnitudes versus redshift for the 80 short and hybrid GRB host galaxies listed in Table \ref{tab:hostlist}. The 14 filled circles represent samples with probable kilonova detection in the light curve, while the other empty circles refer to normal short or hybrid GRB host galaxies. 11 Points overlapping with the thick gray circles are samples with a probability of chance coincidence ($P_{CC}$) measured to be greater than 0.1. The sample marked as an upper limit corresponds to GRB 151229A, whose host galaxy has been measured to have an $i$-band magnitude of 24.924$\pm$0.134 with $P_{CC}$ of 0.04 \citep{2022ApJ...940...56F}.}
\label{fig:sample_rmags}
\end{figure}

With the help of published papers or reports dealing with an individual sGRB and its host galaxy, dozens of host galaxies were added. These cases are GRB 050906, GRB 070810B, GRB 080121, GRB 100216A \citep{2020MNRAS.492.5011D}, GRB 060502B \citep{2007ApJ...654..878B}, GRB 060505 \citep{2007ApJ...662.1129O}, GRB 100816A \citep{2010GCN.11123....1T}, GRB 110402A, GRB 120630A, GRB 160525B \citep{2022MNRAS.515.4890O}, GRB 210510A \citep{2021GCN.30005....1A}, GRB211227A \citep{2022ApJ...931L..23L}, GRB 221120A \citep{2022GCN.32957....1O}. Consequently, our dataset includes 80 short and hybrid GRB host galaxies whose $r$-band magnitude distribution according to their redshift is presented in Figure \ref{fig:sample_rmags}.

\subsection{Ambiguous or Excluded sGRB Host Samples}
Measuring the redshift directly from the faint afterglow of sGRBs is challenging, leading to inherent ambiguities in the identification of their host galaxies. Therefore, if there is no galaxy clearly overlapping in the burst region, we cannot definitively determine whether the host galaxy of such sGRB is a local bright galaxy with a large offset or an undetected faint galaxy at a high redshift.

There are cases where no prominent galaxy can be identified despite the burst locations being accurately determined from optical afterglows. It is worth noting that up to 20\% of sGRBs might occur away from host galaxies with progenitors possibly attaining high velocities of 50-200 km/h due to supernova kicks \citep{1999MNRAS.305..763B, 2006ApJ...648.1110B}. Thus, in cases where no neighboring galaxy is detected in the vicinity (often referred to as ``hostless''), discerning between a high redshift origin and a lower redshift host with significant physical offset becomes particularly complicated \citep{2010ApJ...722.1946B, 2014MNRAS.437.1495T}.

While recent deep and systematic observations have resolved some of these ambiguities, some samples ($\sim$ 13\% of the total samples) remains ambiguous in host identification with the probability of chance coincidence ($P_{CC}$) over 10\%: GRB 050813, GRB 050906, GRB 080121, GRB 080123, GRB 100816A, GRB 110402A, GRB 140622A, GRB 160408A, GRB 200411A, GRB 201221D and GRB 210919A. Note that this probability is obtained from the comparison between the offset with the host and the average number density of galaxies \citep{2002AJ....123.1111B}. These samples could pose a risk of bias, but reports indicate that including such GRB host galaxies does not significantly alter statistical properties \citep{2022ApJ...940...57N}. These samples are still included in our study, and the effects of this will be discussed later.

More such ambiguities arise for events such as GRB 050906, GRB 070810B, GRB 080121, and GRB 100216A, where no afterglow detection exists. While host galaxies are typically not designated without afterglow detection, the $\gamma$-ray error circles for these events overlap with bright galaxies at $<200 \mathrm{Mpc}$ \citep{2020MNRAS.492.5011D}. Such rare coincidences with local bright galaxy fields elevate their significance as host galaxy candidates, though physical associations remain undetermined. In instances of gravitational wave detection, these kinds of galaxies would serve as primary observing targets. We will assess the variance in the GRB host galaxy characteristics with the inclusion or exclusion of these samples to evaluate such prioritization.

Due to insufficient photometric data for the SED modeling, we excluded 5 sGRBs from our sample set: GRB 051210A, GRB 061217, GRB 080905A, GRB 180727A and GRB200907B. We note that our SED modeling necessitates a minimum of five broad-band photometric data points to ensure a robust analysis. Specifically, for GRB 080905A, photometry was compromised by the proximity to bright stars \citep{2013ApJ...769...56F, Nicuesa_Guelbenzu_2021}. These exclusions also can impose a bias on our sample. However, since most are presumed to be at high redshifts, this has a negligible impact on our objective of analyzing the characteristics of nearby GW host galaxies.

\subsection{Afterglow Emission Features and Kilonovae Candidates} \label{sec:kn_sample}

GRBs are classified based on the duration of the prompt emission, i.e., those shorter than 2 seconds are termed sGRBs, while those longer are called long GRBs \citep{1993ApJ...413L.101K}. The criterion of $T_{90}$ less than 2 seconds has been believed to select compact binary coalescence (CBC) among GRBs \citep{1992ApJ...395L..83N}. However, recent studies revealed that collapsars and CBCs significantly overlap in their $T_{90}$ distributions \citep{2013ApJ...764..179B, 2020ApJ...896L..20J}. Given this background, defining a sample based solely on prompt emission might leave contaminants. Therefore, in ambiguous cases, there is a need to investigate the features of the light curve both in prompt and afterglow emissions for further classification.

Among the reported cases, we excluded instances whose $T_{90}$ was shorter than 2 seconds but discovered alongside a supernova, leading to suspicions of being collapsars. GRB 090426 and GRB 200826A are examples of this, and their afterglow brightnesses were consistent with that of typical LGRBs \citep{2009A.A...507L..45A, 2022ApJ...932....1R}. Due to their short delay times, the host galaxies of such collapsars are skewed towards younger stellar age and lower metallicity than regular sGRB host galaxies \citep{2014ARA.A..52...43B}.

Conversely, certain cases warrant classification as CBCs despite having $T_{90}$ duration exceeding 2 seconds. This group notably includes instances of extended emissions (EE), which is characterized by a short, initial peak similar to typical short GRBs, followed by a prolonged tail \citep{2006ApJ...643..266N}. Although the exact physical origin of EE remains uncertain, evidence suggests that their environments, such as offsets and host galaxy luminosities, appear similar to those of regular sGRBs \citep{2022ApJ...940...56F}. Moreover, the $T_{90}$ duration itself is not conclusive as it is subject to instrumental selection effects \citep{2021ApJ...911L..28D}. Consequently, sGRBs exhibiting EE characteristics were included in our sample, regardless of their extended $T_{90}$ duration. The sGRB samples reported to have EE features are indicated in Table~\ref{tab:hostlist}.

Another important aspect is the observation of kilonova (KN) emission features in GRB afterglows. An excess in infrared brightness days after the burst, consistent with the radioactive decay of r-process elements, suggests a significant release of r-process heavy elements. This is indicative of a CBC involving a neutron star. \citet{2019MNRAS.486..672A} identified nine KN candidates through afterglow light curve fitting of historical sGRBs. \citet{2017ApJ...837...50G} classified GRB 050724, GRB 061006, and GRB 070714B as KN candidates based on a magnetar-powered merger-nova model. Additionally, \citet{2020MNRAS.493.3379R} observed signs of shallow decay in the afterglows of six sGRBs (GRB 050709, GRB 050724, GRB 060614, GRB 090515, GRB 150424A, and GRB 160821B) when compared to the light curve of AT2017gfo. Such shallow decays challenge the traditional fireball model, pointing instead towards the presence of an additional emitting source, possibly a kilonova.

Additionally, unique cases have been identified where kilonovae follow typical long GRBs with $T_{90} > 2 \mathrm{sec}$. These events are often referred to as ``hybrid GRBs.'' Notable examples include GRB 060614 \citep{2015NatCo...6.7323Y}, as well as more recent analyses of GRB 060505 \citep{2021arXiv210907694J}, GRB 211211A \citep{2022Natur.612..223R, 2022Natur.612..232Y}, GRB 211227A \citep{2022ApJ...936L..10Z, 2022ApJ...931L..23L}, and GRB 230307A \citep{2023arXiv230900038G, 2024Natur.626..742Y}. The origins of these hybrid GRBs are subject to various interpretations, with some suggesting they might originate from BNS mergers, NSBH mergers, or neutron star-white dwarf mergers \citep{kang2023prospects}. In our study, we include a total of 11 KN candidates and 4 hybrid ones. By analyzing the distribution of KN host galaxies in relation to sGRB host galaxies, we aim to refine our understanding of the environments associated with neutron star mergers.


\begin{longtable*}{cccccccc}
    \caption{List of the Short and Hybrid GRB Host Galaxies.\label{tab:hostlist}} \\
    \toprule
    \multicolumn{1}{c}{GRB} & \multicolumn{1}{c}{redshift} & \multicolumn{1}{c}{$T_{90}$} & \multicolumn{1}{c}{AG detection$^{a}$} & \multicolumn{1}{c}{LC Feature$^{b}$} & \multicolumn{1}{c}{$m_r$} & \multicolumn{1}{c}{Offset} & \multicolumn{1}{c}{$P_{CC}$} \\
    & & \multicolumn{1}{c}{[sec]} & & & \multicolumn{1}{c}{[ABmag]} & \multicolumn{1}{c}{[arcsec]} & \\
    \midrule
    \endfirsthead
    
    \multicolumn{8}{c}%
    {{\tablename\ \thetable{} -- continued from previous page}} \\
    \toprule
    \multicolumn{1}{c}{GRB} & \multicolumn{1}{c}{redshift} & \multicolumn{1}{c}{$T_{90}$} & \multicolumn{1}{c}{AG detection$^{a}$} & \multicolumn{1}{c}{LC Feature$^{b}$} & \multicolumn{1}{c}{$m_r$} & \multicolumn{1}{c}{Offset} & \multicolumn{1}{c}{$P_{CC}$} \\
    & & \multicolumn{1}{c}{[sec]} & & & \multicolumn{1}{c}{[ABmag]} & \multicolumn{1}{c}{[arcsec]} & \\
    \midrule
    \endhead
    
    \midrule
    \multicolumn{8}{r}{{Continued on next page}} \\
    \midrule
    \endfoot
    
    \bottomrule
    \endlastfoot
    
GRB050509B &                  $0.225$ &   0.024 &     X-ray &           - &  17.30±0.07 &   15.10 & 5.00E-03 \\
 GRB050709 &                  $0.161$ &   0.070 &   Optical &          KN$^{1, 2}$ &  21.23±0.07 &    1.35 & 3.00E-03 \\
 GRB050724 &                  $0.254$ &  98.0 &   Optical &      KN$^{1, 2, 3}$, EE &  18.81±0.05 &    0.68 & 2.00E-05 \\
 GRB050813 &                  $0.719$ &   0.380 &     X-ray &           - &  23.43±0.07 &    5.96 & 0.2 \\
 GRB050906 &                 $0.0308$ &   0.128 & None &           - &  13.78±0.03 &  162.63 & 0.144 \\
 GRB051210 &  $0.681^{+0.27}_{-0.34}$ &   1.30 &     X-ray &           - &  24.04±0.15 &    3.56 & 0.04 \\
GRB051221A &                  $0.546$ &   1.40 &   Optical &           - &  22.67±0.09 &    0.32 & 5.00E-05 \\
GRB060502B &                  $0.287$ &   0.131 &     X-ray &           - &  18.95±0.04 &   15.91 & 0.076 \\
 GRB060505 &                 $0.0894$ &   4.00 &   Optical &      KN$^{4}$, EE &  18.05±0.04 &    3.60 & 2.03E-03 \\
 GRB060614 &                  $0.125$ & 109 &   Optical &      KN$^{1, 2, 5}$, EE &  22.60±0.05 &    0.31 & 3.00E-04 \\
 GRB060801 &                   $1.13$ &   0.490 &   Optical &           - &  23.15±0.11 &    1.23 & 0.02 \\
 GRB061006 &                  $0.461$ & 130 &   Optical &      KN$^{3}$, EE &  23.28±0.09 &    0.24 & 4.00E-04 \\
 GRB061210 &                   $0.41$ &  85.3 &     X-ray &          EE &  21.71±0.13 &    2.82 & 0.02 \\
GRB070429B &                  $0.902$ &   0.470 &   Optical &           - &  23.21±0.04 &    0.76 & 3.00E-03 \\
GRB070714B &                  $0.923$ &  64.0 &   Optical &      KN$^{3}$, EE &  24.54±0.22 &    1.55 & 5.00E-03 \\
 GRB070724 &                  $0.457$ &   0.400 &   Optical &           - &  20.72±0.05 &    0.94 & 8.00E-04 \\
 GRB070729 &  $0.761^{+0.08}_{-0.08}$ &   0.900 &     X-ray &           - &  22.94±0.09 &    3.13 & 0.036 \\
 GRB070809 &                  $0.473$ &   1.30 &   Optical &          KN$^{6}$ &  19.89±0.02 &    5.70 & 6.00E-03 \\
GRB070810B &                 $0.0385$ &   0.072 & None &           - &  14.66±0.01 &   81.04 & 0.073 \\
 GRB071227 &                  $0.381$ & 143 &   Optical &          EE &  20.72±0.03 &    2.80 & 0.01 \\
 GRB080121 &                  $0.045$ &   0.320 & None &           - &  17.68±0.05 &   66.96 & 0.41 \\
 GRB080123 &                  $0.495$ & 115 &     X-ray &          EE &  21.04±0.04 &    8.74 & 0.11 \\
 GRB090510 &                  $0.903$ &   5.66 &   Optical &          EE &  22.41±0.14 &    1.33 & 8.00E-03 \\
 GRB090515 &                  $0.403$ &   0.036 &     X-ray &       EE &  20.21±0.05 &   13.98 & 0.05 \\
 GRB100117 &                  $0.914$ &   0.300 &   Optical &          EE &  24.33±0.10 &    0.17 & 7.00E-05 \\
GRB100206A &                  $0.407$ &   0.120 &     X-ray &           - &  20.55±0.09 &    4.59 & 0.02 \\
GRB100216A &                  $0.038$ &   0.208 & None &           - &  16.71±0.07 &   34.66 & 0.065 \\
GRB100625A &                  $0.452$ &   0.330 &     X-ray &           - &  22.61±0.05 &    0.45 & 0.04 \\
GRB100816A &                  $0.805$ &   2.05 &   Optical &           - &  21.89±0.06 &    6.39 & 0.115 \\
GRB101219A &                  $0.718$ &   0.830 &     X-ray &           - &  23.95±0.05 &    0.75 & 0.06 \\
GRB101224A &                  $0.454$ &   0.200 &     X-ray &           - &  22.07±0.05 &    2.18 & 0.015 \\
GRB110402A &                  $0.854$ &  56.0 &   Optical &          EE &  24.24±0.20 &    3.39 & 0.19 \\
GRB111117A &                   $2.21$ &   0.470 &     X-ray &           - &  23.86±0.08 &    1.25 & 0.024 \\
GRB120305A &                  $0.225$ &   0.100 &     X-ray &           - &  22.40±0.05 &    4.97 & 0.053 \\
GRB120630A &  $0.544^{+0.13}_{-0.08}$ &   0.600 &   Optical &           - &  21.60±0.06 &    3.39 & 0.027 \\
GRB120804A &  $1.258^{+0.36}_{-0.37}$ &   0.810 &   Optical &           - &  26.41±0.20 &    0.27 & 0.02 \\
GRB121226A &  $2.103^{+0.20}_{-0.18}$ &   1.00 &     X-ray &           - &  24.31±0.06 &    0.27 & 0.019 \\
GRB130515A &  $0.891^{+0.05}_{-0.05}$ &   0.290 &     X-ray &           - &  22.65±0.04 &    8.05 & 0.081 \\
GRB130603B &                  $0.357$ &   0.180 &   Optical &          KN$^{1, 7}$ &  20.82±0.01 &    1.07 & 2.00E-03 \\
GRB130822A &                  $0.154$ &   0.040 &     X-ray &           - &  18.25±0.06 &   22.32 & 0.086 \\
GRB140129B &                   $0.43$ &   1.36 &   Optical &           - &  23.55±0.07 &    0.31 & 8.70E-04 \\
GRB140622A &                  $0.959$ &   0.130 &     X-ray &           - &  22.70±0.04 &    4.10 & 0.1 \\
GRB140903A &                  $0.353$ &   0.300 &   Optical &           - &  20.51±0.09 &    0.18 & 6.20E-05 \\
GRB140930B &                  $1.465$ &   0.840 &   Optical &           - &  24.21±0.25 &    1.12 & 0.021 \\
GRB141212A &                  $0.596$ &   0.300 &     X-ray &           - &  22.95±0.06 &    2.78 & 2.90E-04 \\
GRB150101B &                  $0.134$ &   0.018 &   Optical &          KN$^{1, 8}$ &  16.56±0.04 &    3.07 & 4.80E-04 \\
GRB150120A &                   $0.46$ &   1.20 &     X-ray &          EE &  22.05±0.06 &    0.81 & 1.90E-03 \\
GRB150728A &                  $0.461$ &   0.830 &     X-ray &           - &  21.42±0.05 &    1.28 & 0.018 \\
GRB150831A &                   $1.18$ &   0.920 &     X-ray &           - &  24.43±0.45 &    1.48 & 0.037 \\
GRB151229A &  $1.879^{+0.80}_{-0.80}$ &   1.44 &     X-ray &           - & $\gtrsim$24.49 &    1.18 & 0.04 \\
GRB160303A &  $1.095^{+0.28}_{-0.28}$ &   5.00 &   Optical &           - &  25.80±0.30 &    1.88 & 0.096 \\
GRB160408A &  $1.483^{+0.72}_{-0.65}$ &   0.320 &   Optical &           - &  25.74±0.16 &    1.65 & 0.14 \\
GRB160411A &  $0.732^{+0.37}_{-0.42}$ &   0.360 &     X-ray &           - &  24.58±0.13 &    0.18 & 7.20E-04 \\
GRB160525B &  $0.576^{+0.07}_{-0.08}$ &   0.290 &     X-ray &           - &  23.28±0.09 &    0.79 & 1.00E-03 \\
GRB160624A &                  $0.484$ &   0.200 &     X-ray &          EE &  22.03±0.02 &    1.59 & 0.037 \\
GRB160821B &                  $0.162$ &   0.480 &   Optical &      KN$^{1, 2, 9}$, EE &  19.49±0.07 &    5.61 & 0.044 \\
GRB161001A &  $0.776^{+0.07}_{-0.07}$ &   2.60 &     X-ray &           - &  22.97±0.05 &    2.61 & 0.045 \\
GRB161104A &                  $0.793$ &   0.100 &     X-ray &           - &  23.43±0.10 &    0.22 & 0.06 \\
GRB170127B &  $1.443^{+0.42}_{-0.33}$ &   0.510 &     X-ray &           - &  25.32±0.29 &    1.24 & 0.098 \\
GRB170428A &                  $0.453$ &   0.200 &   Optical &           - &  22.35±0.10 &    1.32 & 6.70E-03 \\
GRB170728A &                  $1.493$ &   1.25 &     X-ray &           - &  24.73±0.14 &    3.75 & 0.22 \\
GRB170728B &                  $1.272$ &  47.7 &   Optical &           - &  23.31±0.10 &    0.99 & 8.30E-03 \\
 GRB170817 &                $0.00979$ &   2.64 &   Optical &          KN$^{1, 10, 11}$ &  12.16±0.01 &   10.31 & 4.90E-04 \\
GRB180418A &  $1.094^{+0.78}_{-0.63}$ &   2.29 &   Optical &          EE &  25.69±0.21 &    0.16 & 1.50E-03 \\
GRB180618A &  $0.643^{+0.09}_{-0.09}$ &  47.4 &   Optical &          EE &  23.47±0.13 &    1.54 & 8.20E-03 \\
GRB180805B &                  $0.661$ & 123 &     X-ray &          EE &  22.15±0.06 &    3.44 & 0.042 \\
GRB181123B &                   $1.75$ &   0.260 &   Optical &           - &  23.84±0.19 &    0.59 & 4.40E-03 \\
GRB191031D &  $1.607^{+0.33}_{-0.24}$ &   0.290 &     X-ray &           - &  24.46±0.26 &    1.53 & 0.043 \\
GRB200219A &  $0.457^{+0.18}_{-0.21}$ & 288 &     X-ray &          EE &  20.66±0.05 &    1.38 & 2.20E-03 \\
GRB200411A &  $1.145^{+0.07}_{-0.06}$ &   0.220 &     X-ray &           - &  22.56±0.04 &    4.91 & 0.11 \\
GRB200522A &                  $0.554$ &   0.620 &   Optical &          KN$^{12}$ &  21.28±0.05 &    0.14 & 3.50E-05 \\
GRB201221D &                  $1.055$ &   0.160 &   Optical &           - &  23.42±0.08 &    3.57 & 0.12 \\
GRB210323A &                  $0.733$ &   1.12 &   Optical &           - &  24.97±0.25 &    0.80 & 0.013 \\
GRB210510A &                  $0.221$ &   1.34 &   Optical &           - &  21.03±0.09 &    1.34 & 2.79E-03 \\
GRB210726A & $0.458^{+0.17}_{-0.18}$ &   0.390 &     X-ray &           - &  22.16±0.09 &    0.04 & 7.30E-05 \\
GRB210919A &                  $0.242$ &   0.160 &     X-ray &           - &  20.50±0.05 &   13.28 & 0.13 \\
GRB211023B &                  $0.862$ &   1.30 &   Optical &           - &  24.36±0.38 &    0.49 & 4.70E-03 \\
GRB211211A &                 $0.0763$ &  51.4 &   Optical &          KN$^{13, 14}$ &  19.46±0.04 &    5.44 & 0.013 \\
GRB211227A &                  $0.228$ &  83.8 &     X-ray &          KN$^{15, 16}$ &  19.26±0.05 &    0.38 & 5.76E-05 \\
GRB221120A &   $0.08^{+0.04}_{-0.03}$ &   0.640 &   Optical &           - &  17.53±0.05 &   15.66 & 0.025 \\
\midrule
\multicolumn{8}{l}{\textbf{Notes:}} \\
\multicolumn{8}{p{6in}}{$^{a}$ Afterglow detection methods. `Optical' and `X-ray' indicate afterglows detected by \emph{Swift} UVOT and XRT, respectively. `None' represents samples only detected by $\gamma$-ray, but coincide with local bright galaxies.} \\
\multicolumn{8}{p{6in}}{$^{b}$ Light curve emission features. `KN' indicates possible kilonova emission included in the light curve, and `EE' represents the extended emission feature in the prompt emission. \textbf{Reference} $^{1}$\citet{2019MNRAS.486..672A}, $^{2}$\citet{2020MNRAS.493.3379R}, $^{3}$\citet{2017ApJ...837...50G},  $^{4}$\citet{2021arXiv210907694J}, $^{5}$\citet{2015NatCo...6.7323Y}, $^{6}$\citet{2020NatAs...4...77J}, $^{7}$\citet{2013Natur.500..547T}, $^{8}$\citet{2018NatCo...9.4089T}, $^{9}$\citet{2019MNRAS.489.2104T}, $^{10}$\citet{2017ApJ...848L..17C}, $^{11}$\citet{2017ApJ...848L..27T}, $^{12}$\citet{2021ApJ...906..127F}, $^{13}$\citet{2022Natur.612..223R}, $^{14}$\citet{2022Natur.612..232Y}, $^{15}$\citet{2022ApJ...936L..10Z}, $^{16}$\citet{2022ApJ...931L..23L}.} \\
\end{longtable*}

\section{Photometric Data Collection} \label{sec:data}

All the data used in this study are derived either from archival image data or from publicly available observational studies. For each host galaxy, direct photometry was performed wherever images were available, while for others, published photometry information was utilized. We ensured consistent magnitude extraction from these varied sources by carefully adjusting the photometric apertures. Additionally, we gathered filter curves corresponding to each instrument and conducted extinction corrections uniformly. This photometric data, in conjunction with redshift information, are instrumental for the SED fitting discussed in the following section.

\subsection{Imaging Survey Data}

Large-area survey data enable the collection of photometric information for host galaxies at arbitrary coordinates. This approach allows for the utilization of accurately calibrated multi-wavelength photometric data. Additionally, survey image data are free from contamination by transients, providing reliable magnitudes. Despite limitations in depth, we endeavored to collect extensive photometric information from brighter galaxies.

For optical data, we queried images from the Pan-STARRS (PS1) DR2 \citep{2012ApJ...750...99T, 2016arXiv161205560C} and the Sloan Digital Sky Survey (SDSS) DR12 \citep{2015ApJS..219...12A}. For targets in the southern hemisphere, images from the Dark Energy Survey (DES) DR2 \citep{2021ApJS..255...20A} and the Dark Energy Camera Legacy Survey (DECaLS) DR8 \citep{2019AJ....157..168D} were used. 

UV and NIR survey data were also incorporated. The Galactic Evolution Explorer (GALEX) provided $NUV$-band data \citep{2005ApJ...619L...1M}, while the Wide-field Infrared Survey Explorer (WISE) \citep{2010AJ....140.1868W}, UKIRT Hemisphere Survey (UHS) \citep{2018MNRAS.473.5113D}, and UKIRT Infrared Deep Sky Survey (UKIDSS) \citep{2007MNRAS.379.1599L, 2012yCat.2314....0L} supplied NIR magnitudes. These values were directly referenced from their respective source catalogs. Additional data were accessed through the Gemini Observatory Archive and the Mikulski Archive for Space Telescopes (MAST), with detailed sources recorded in Table \ref{tab:hostphot}.

In photometry, \texttt{Source Extractor} was used, adopting \texttt{AUTO} apertures to estimate total magnitudes for extended sources \citep{1996A.AS..117..393B}. We utilized \texttt{MAG\_AUTO} for each band image as the photometric value, instead of unifying all aperture sizes with dual-mode photometry. Regarding the magnitude zero-point calibration of PS1 images, field stars in the PS1 catalog were used as secondary standard stars. For DES and DECaLS data, standardized zero-points of 30.0 ABmag and 22.5 ABmag were adopted, respectively.

Finally, galactic extinction correction was performed using E(B-V) estimates from \citealt{2011ApJ...737..103S} (SF11), a recalibration of the dust map by \citealt{1998ApJ...500..525S} (SFD98). Extinction values were converted using the curve from \citet{1999PASP..111...63F}, adopting $R_{V}$=3.1.

\subsection{Data in Literature}

We collected magnitudes of host galaxies published in previous studies to supplement our photometric data. Notably, several population studies have targeted sGRB host galaxies. \citet{2010ApJ...725.1202L} compiled optical/NIR photometric data for 19 host galaxies. Additionally, \citet{2013ApJ...769...56F} provided in-depth photometric information for 10 sGRB host galaxies using the Hubble Space Telescope (HST). A detailed review of 39 known sGRB host galaxies was conducted by \citet{2014ARA.A..52...43B}.

Recent deep survey observations have significantly expanded the catalog of known sGRB host galaxies. \citet{2022MNRAS.515.4890O} used the 4.3m Lowell Discovery Telescope for deep observations (to $r\gtrsim$ 25 ABmag) in fields of X-ray or optical sGRB afterglows, where the host galaxies were previously unknown. In cases with unclear coincident host galaxies, further in-depth observations ($r\gtrsim$ 26-28 ABmag) were carried out using the Gemini and Keck-I telescopes. Systematic observations ($r\simeq$ 24-26 ABmag) of all 90 sGRBs with afterglows discovered by \textit{Swift} from 2005 to 2021 were performed by \citet{2022ApJ...940...56F}, providing multi-band photometry for 84 of them and spectral data for 25 bright galaxies. For the rest of 6 sGRBs, no definitive host galaxy was identified despite optical/NIR observations reaching depths of 26-28 ABmag.

The comprehensive and systematic efforts of these studies have yielded extensive optical and NIR band photometric data. Some corrections, however, were necessary in specific cases. For instance, the host galaxy of GRB 161104A, situated in a cluster, was studied by \citet{2020ApJ...904...52N} using a smaller aperture to mitigate nearby galaxy effects. This approach may underestimate the galaxy's stellar mass compared to others in the SED modeling. Thus, we re-analyzed the $r$-band image from the Gemini Observatory Archive and calculated \texttt{MAG\_AUTO}. We obtained the brightness difference of 0.38 ABmag compared to \citet{2020ApJ...904...52N}, which was considered as an aperture correction factor and was applied to other bands.

Another case warranting careful examination involved the host of GRB 051210, for which the $r$ and $i$ band fluxes demonstrated inadequate fits in their SED, as documented in \citep{2010ApJ...725.1202L}. An analysis of the DES data indicated that the $i$-band magnitude should be adjusted to be approximately 0.5 mag brighter. Furthermore, the $F675W=21.14 \pm 0.05$ reported in \citep{2010ApJ...708....9F} was deemed potentially inaccurate due to its higher brightness in comparison to the $r$-band magnitude. Instead, $F675W=23.7$, derived through \texttt{GALFIT} modeling in the same study, was adopted. By revising the magnitudes for the fitting process, we considered this galaxy to be a comparatively less massive galaxy at $z \sim 0.6$, as opposed to being a higher-redshift galaxy as initially conjectured. Due to the absence of spectroscopic redshift data, the estimated properties of this host galaxy remain relatively uncertain. The data referenced herein and findings from previous surveys are listed in Table \ref{tab:hostphot} in the appendix.

\subsection{Galactic Extinction Correction}

One of the conscious tasks of unifying photometric data from various references is the galactic extinction correction. To determine the amount of dust, the SFD98 dust map used IRAS and COBE all-sky observation data, while recent studies have often used SF11, which conducted additional calibration to the SFD98 work.

The correction values of SF11 are approximately 14\% lower than those of SFD98. To maintain consistency, we recalibrated the magnitudes from referenced papers that used SFD as the basis for extinction correction. We queried the values of E(B-V) at the locations of each host galaxy from the extinction calculator provided by the NASA Extragalactic Database (NED). This allowed us to accurately correct for E(B-V) because extinction values for the Landolt B and V filters are given according to the methods of SFD98 and SF11. While some studies have referred \citet{1989ApJ...345..245C} for the extinction law, we regarded the uncertainty arising from this difference as negligible.

\subsection{Filter Transmission Curves}

We are utilizing photometric values collected from various surveys and independent papers, so it is necessary to reflect the filter transmission curves suitable for each instrument to accurately measure the flux density at each wavelength. The Spanish Virtual Observatory (SVO) filter profile service provides filter curves for various instruments, making it convenient to utilize the filter information used in this study \citep{2020sea..confE.182R}. For the filter profiles that are not available in the SVO service, we adopted the profiles provided by each respective observatory. Details of the instruments, filter, and corresponding central wavelengths can be found in the Table \ref{tab:hostphot} in the appendix.

\section{Host Galaxy Properties} \label{sec:sedmodelling}

In this section, we outline the method of estimating the properties of host galaxies. Utilizing the photometric data and spectroscopic redshift, the best-fit stellar population model was determined, providing information on galaxies' various properties. Additionally, we applied a similar approach to the control sample galaxies for comparative analysis.

\begin{figure*}[t]
\begin{center}
\includegraphics[width=\linewidth]{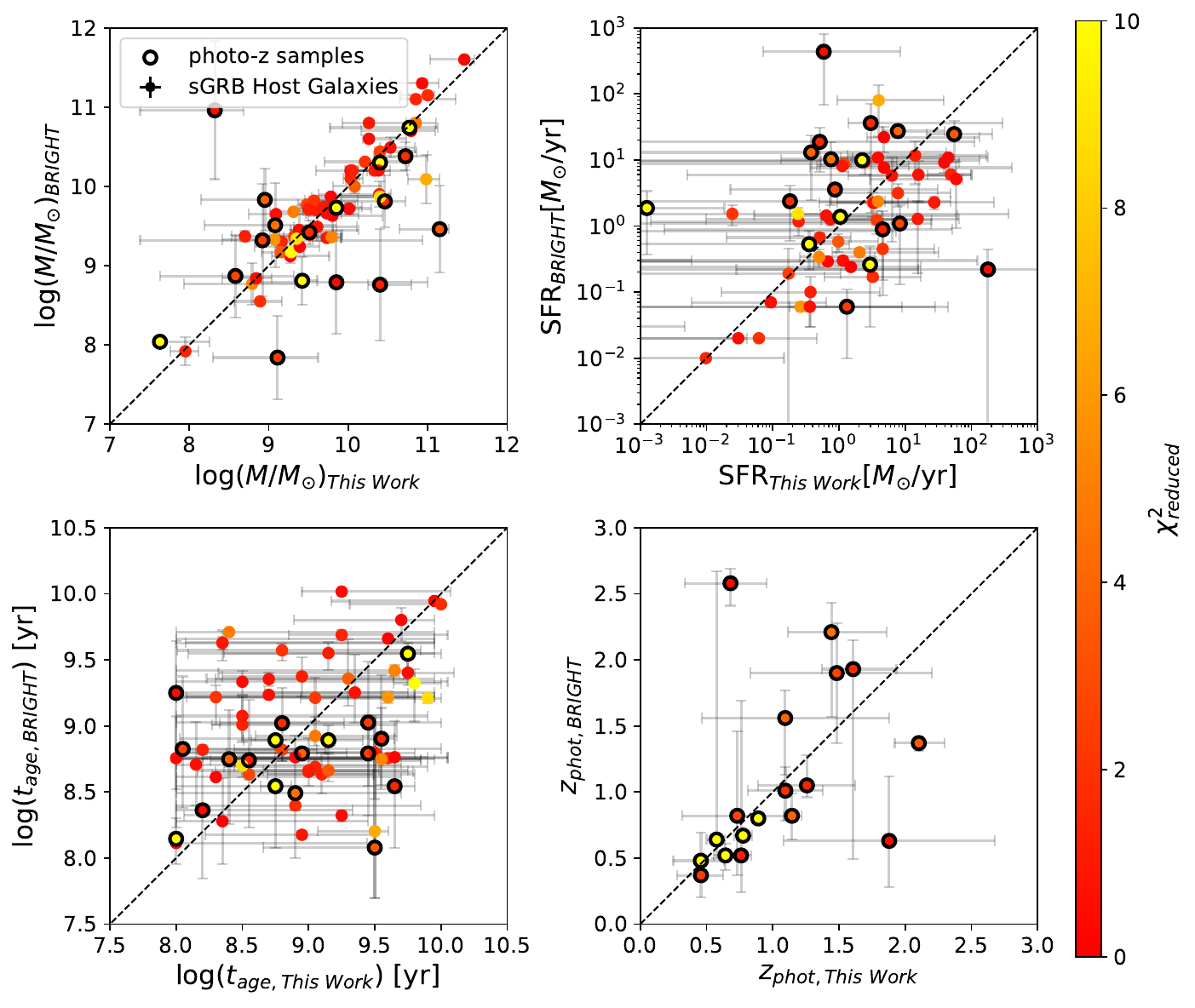}

\caption{A comparison of host galaxy properties derived from our SED fitting and photometric redshifts with those from \citet{2022ApJ...940...57N}. The comparison is made for 67 galaxies that are cataloged in both studies. The properties compared are the stellar mass (\textbf{top-left}), the star formation rate (\textbf{top-right}), the stellar population age (\textbf{bottom-left}), and the photometric redshift (\textbf{bottom-right}). The data points with black outlines indicate the host galaxy samples lacking spectroscopic redshift data. In each panel, the x-axis represents values reported in this work, while the y-axis corresponds to those reported by \citet{2022ApJ...940...57N}. The colored scale reflects the reduced chi-squared values from the SED fitting.
\label{fig:refcomp}}
\end{center}
\end{figure*}
\subsection{Photometric Redshift Estimation}

Before conducting SED fitting, it was necessary to estimate photometric redshifts (photo-z) for targets lacking spectroscopic data. Among the 80 samples, 20 galaxies required photo-z measurements, each with more than five broad-band photometry data points. We used the software Easy and Accurate Redshifts from Yale \citep[\texttt{EAzY};][]{2008ApJ...686.1503B} for this task

The magnitudes and filter information were input into \texttt{EAzY}, and the data points were compared to synthetic models from \citet{2003MNRAS.344.1000B} to determine the best-fit model via $\chi^2$ minimization. The redshift range for fitting extended from 0 to 3, with each step size set at 0.01$\times$(1+z). 

It is important to note that photo-z estimates can sometimes exhibit color degeneracy across different redshift ranges. Following the methodology of \citet{2008ApJ...686.1503B}, we used the $r$-band apparent magnitude as a prior, allowing for more reasonable photo-z estimates by considering luminosity function in a cosmic volume.

In alignment with our objective to study BNS mergers in the nearby universe, our analysis was limited to host galaxies at $z<0.5$. We determined the representative photo-z as the $z_{m2}$ parameter of \texttt{EAzY} output. The $z_{m2}$ represent the redshift marginalized over the posterior distribution, which is generally considered the best photo-z estimate. Within this range, we identified three host galaxies with photo-z are suitable for inclusion to the $z<0.5$ sample: GRB 200219A ($z_{phot}=0.457^{+0.177}_{-0.208}$), GRB 210726A ($z_{phot}=0.458^{+0.170}_{-0.181}$), and GRB 221120A ($z_{phot}=0.08^{+0.036}_{-0.032}$). When comparing our photo-z estimates to those derived in \citet{2022ApJ...940...57N}, we find a generally reasonable agreement within measurement errors (Figure \ref{fig:refcomp}). Notably, there are no instances where the photo-z of a host galaxy was estimated to be less than 0.5 in other studies but excluded from our criteria.

GRB 051210 ($z_{phot}=0.681^{+0.273}_{-0.344}$) and GRB 151229A ($z_{phot}=1.879^{+0.796}_{-0.799}$) show significant discrepancies compared to the literature values. For GRB 051210, this difference arises from updates to the photometric data. We found that the SED model presented by \citet{2010ApJ...725.1202L} did not adequately explain the observations in the $r$ and $i$ bands. To improve the model, we incorporated $i$-band data from DES and $F675W$ magnitude from \citet{2010ApJ...708....9F}. This resulted in a more reasonable SED model and a lower photo-z estimate. In contrast, for GRB 151229A, the absence of detections in the $g$ and $r$ bands makes photo-z measurements highly uncertain for both our estimates and those in the literature.

We also note that the posterior distribution of GRB 200219A is bimodal with the redshift peaks at $z=0.28$ and $0.6$. We classified this sample as $z<0.5$ based on the estimate of $z_{m2} = 0.458$, but the exclusion of this object from the $z<0.5$ sample or adaption of the $z=0.28$ value for the redshift should not significantly affect our analysis results.

\subsection{SED Modelling}

Estimating the spectral energy distribution (SED) using multi-wavelength data of galaxies is an effective approach to inferring their global properties. Although SED fitting results are available for most of our sGRB host galaxies, they are collected from various sources that used different SED fitting methods which could create a bias in subsequent analysis results. Hence, we decided to perform the SED fitting for all of our sample host galaxies adopting a single, consistent method. We employed the Fitting and Assessment of Synthetic Templates (\texttt{FAST}; \citep{2009ApJ...700..221K}) program, which interworks with the photo-z outputs from \texttt{EAzY}.

We adopted the stellar population library from \citealt{2003MNRAS.344.1000B} (BC03) and assumed the initial mass function of \citet{2003PASP..115..763C}. The dust attenuation law was based on \citet{2000ApJ...533..682C}. We assumed a delayed exponential form for the star formation history (SFH), represented as SFR $\propto t \times \exp(-t/\tau)$, where $\tau$ is the e-folding time scale of SFR.

The output parameter space was established as a grid. Specifically, we varied log($\tau$/yr) in 0.1 steps within the range [7.6, 10.0]. We also varied log($t_{age}$/yr), the mean stellar population age, in 0.1 steps within the range [8.0, 10.1].  To ensure that the stellar population age does not exceed the age of the Universe at each redshift of the host galaxy, we set the maximum age limit to be the age of the Universe at that specific redshift. $A_{V}$ was varied in units of 0.1mag within the range of [0, 4.7], and three values were adopted for the metallicity parameter $Z/Z_{\odot}$ as subsolar (0.004, 0.008), solar (0.02) and supersolar (0.05).

We determined the best-fit model for the given flux and redshift through the aforementioned process. Each fitting process involved 100 Monte Carlo simulations within the error range of the input flux, from which the best-fit parameters and their 1 $\sigma$ errors were estimated. The results are tabulated in Table \ref{tab:fitresult}, with each column representing the redshift, the stellar mass, the star formation rate, the stellar age, the metallicity, the attenuation, and the reduced $\chi^2$, respectively.

In Figure \ref{fig:refcomp}, we compared the stellar mass, the star formation rate, and the mean stellar population age values from our SED fitting with the results from previous studies \citep{2022ApJ...940...57N}. While we used the same initial mass function \citep{2003PASP..115..763C} and the star formation history (SFH) model, \citet{2022ApJ...940...57N} adopted the Flexible Stellar Population Synthesis (FSPS) as the stellar population model and the Milky Way extinction law for the host internal extinction \citep{2009ApJ...699..486C, 1989ApJ...345..245C}. 

The stellar mass values from the two studies are almost consistent within the error range, except for a few galaxies with inaccurate photo-z estimates. For example, the $i$-band magnitude of the GRB 051210 host galaxy, as explained in Section~\ref{sec:data}, was updated by us, lowering the estimated stellar mass value by about two orders of magnitude compared to the value from \citet{2022ApJ...940...57N}. SFR estimates from the two studies are generally in agreement with each other within errors but with relatively large errors compared to the stellar masses. The large errors in mean stellar age comparisons highlighted the challenges in precise age determination due to the color and metallicity degeneracy.

\begin{longtable*}{lllllllllr}
    \caption{The Best-fit Parameters of the GRB Host Galaxies \label{tab:fitresult}} \\
    \toprule
    \multicolumn{1}{c}{GRB} & \multicolumn{1}{c}{redshift} & \multicolumn{1}{c}{log($M_{*}$)} & \multicolumn{1}{c}{log(SFR)} & \multicolumn{1}{c}{$\log(\mathrm{sSFR})$} & \multicolumn{1}{c}{Type$^{a}$} & \multicolumn{1}{c}{$t_{age}$} & \multicolumn{1}{c}{$Z$} & \multicolumn{1}{c}{$A_{V}$} & \multicolumn{1}{c}{$\chi^{2}_{reduced}$} \\
    & & \multicolumn{1}{c}{[$M_{\odot}$]} & \multicolumn{1}{c}{[$M_{\odot}$/yr]} & \multicolumn{1}{c}{[yr$^{-1}$]} & & \multicolumn{1}{c}{[Gyr]} & \multicolumn{1}{c}{[$Z_{\odot}$]} & \multicolumn{1}{c}{[mag]} & \\
    \midrule
    \endfirsthead
    
    \toprule
    \multicolumn{1}{c}{GRB} & \multicolumn{1}{c}{redshift} & \multicolumn{1}{c}{log($M_{*}$)} & \multicolumn{1}{c}{log(SFR)} & \multicolumn{1}{c}{$\log(\mathrm{sSFR})$} & \multicolumn{1}{c}{Type$^{a}$} & \multicolumn{1}{c}{$t_{age}$} & \multicolumn{1}{c}{$Z$} & \multicolumn{1}{c}{$A_{V}$} & \multicolumn{1}{c}{$\chi^{2}_{reduced}$} \\
    & & \multicolumn{1}{c}{[$M_{\odot}$]} & \multicolumn{1}{c}{[$M_{\odot}$/yr]} & \multicolumn{1}{c}{[yr$^{-1}$]} & & \multicolumn{1}{c}{[Gyr]} & \multicolumn{1}{c}{[$Z_{\odot}$]} & \multicolumn{1}{c}{[mag]} & \\
    \midrule
    \endhead
    
    \midrule
    \multicolumn{10}{r}{{Continued on next page}} \\
    \midrule
    \endfoot
    
    \bottomrule
    \endlastfoot

GRB050509B &                 $0.225$ & $11.46^{+0.11}_{-0.39}$ &   $0.42^{+0.13}_{-0.00}$ & $-11.04^{+0.20}_{-0.00}$ &         Q &  $8.91^{+1.09}_{-7.33}$ & $0.004^{+0.02}_{-0.00}$ & $0.4^{+0.55}_{-0.40}$ &        0.24 \\
 GRB050709 &                 $0.161$ &  $8.89^{+0.27}_{-0.33}$ &  $-1.21^{+0.87}_{-0.97}$ &  $-10.1^{+1.05}_{-0.97}$ &        SF &  $3.98^{+7.24}_{-3.71}$ & $0.004^{+0.03}_{-0.00}$ & $0.2^{+1.03}_{-0.20}$ &        1.76 \\
 GRB050724 &                 $0.254$ &  $11.0^{+0.14}_{-0.87}$ &   $0.51^{+2.08}_{-0.54}$ & $-10.48^{+2.86}_{-0.55}$ &        SF &  $10.0^{+0.00}_{-9.90}$ &  $0.05^{+0.00}_{-0.05}$ & $0.2^{+2.58}_{-0.20}$ &        1.67 \\
 GRB050813 &                 $0.719$ & $10.21^{+0.59}_{-0.51}$ &  $-0.65^{+3.36}_{-0.00}$ & $-10.86^{+2.93}_{-0.00}$ &         Q &  $0.63^{+6.45}_{-0.53}$ & $0.008^{+0.04}_{-0.00}$ & $1.4^{+2.65}_{-1.40}$ &        1.97 \\
 GRB050906 &                $0.0308$ & $10.23^{+0.33}_{-0.45}$ &   $-0.9^{+2.71}_{-0.00}$ & $-11.13^{+3.16}_{-0.00}$ &         Q & $0.71^{+11.88}_{-0.58}$ &  $0.02^{+0.02}_{-0.02}$ & $0.4^{+0.98}_{-0.40}$ &        0.63 \\
 GRB051210 & $0.681^{+0.27}_{-0.34}$ &  $8.32^{+0.39}_{-0.98}$ &  $-0.23^{+1.18}_{-1.14}$ &  $-8.55^{+0.94}_{-0.58}$ &        SF &  $0.16^{+2.24}_{-0.06}$ &  $0.05^{+0.00}_{-0.05}$ & $0.0^{+1.14}_{-0.00}$ &        0.41 \\
GRB051221A &                 $0.546$ &  $9.16^{+0.23}_{-0.19}$ &  $-0.29^{+0.42}_{-1.07}$ &  $-9.45^{+0.50}_{-0.96}$ &        SF &  $1.12^{+6.29}_{-0.88}$ & $0.004^{+0.01}_{-0.00}$ & $0.0^{+0.44}_{-0.00}$ &        1.36 \\
GRB060502B &                 $0.287$ & $11.08^{+0.12}_{-0.56}$ &   $0.06^{+0.58}_{-0.00}$ & $-11.02^{+0.70}_{-0.00}$ &         Q &  $10.0^{+0.00}_{-9.15}$ & $0.008^{+0.04}_{-0.00}$ & $0.1^{+1.33}_{-0.10}$ &        0.39 \\
 GRB060505 &                $0.0894$ &  $9.33^{+0.38}_{-0.41}$ &    $0.4^{+0.97}_{-0.70}$ &  $-8.93^{+1.31}_{-0.96}$ &        SF & $1.78^{+10.81}_{-1.68}$ & $0.004^{+0.05}_{-0.00}$ & $0.6^{+0.90}_{-0.60}$ &        1.17 \\
 GRB060614 &                 $0.125$ &  $7.95^{+0.13}_{-0.19}$ &  $-2.01^{+1.14}_{-1.12}$ &  $-9.96^{+1.19}_{-1.12}$ &        SF &  $0.32^{+3.31}_{-0.13}$ &  $0.05^{+0.00}_{-0.05}$ & $0.8^{+0.64}_{-0.80}$ &        1.44 \\
 GRB060801 &                  $1.13$ &  $9.27^{+0.39}_{-0.21}$ &   $1.59^{+0.22}_{-1.07}$ &  $-7.68^{+0.07}_{-1.13}$ &        SF &   $0.1^{+1.56}_{-0.00}$ &  $0.05^{+0.00}_{-0.05}$ & $1.0^{+0.34}_{-1.00}$ &        0.14 \\
 GRB061006 &                 $0.461$ &   $8.7^{+0.47}_{-0.17}$ &  $-0.43^{+1.07}_{-1.34}$ &  $-9.13^{+1.14}_{-1.19}$ &        SF &  $0.22^{+8.69}_{-0.11}$ & $0.008^{+0.04}_{-0.00}$ & $0.7^{+0.71}_{-0.70}$ &        1.15 \\
 GRB061210 &                  $0.41$ &  $9.48^{+0.11}_{-0.11}$ &  $-0.76^{+0.90}_{-0.73}$ & $-10.24^{+0.87}_{-0.73}$ &        SF &  $0.63^{+0.85}_{-0.36}$ &  $0.05^{+0.00}_{-0.03}$ & $0.3^{+0.81}_{-0.30}$ &        2.45 \\
GRB070429B &                 $0.902$ &  $10.4^{+0.42}_{-0.07}$ &   $0.08^{+1.45}_{-0.48}$ & $-10.32^{+1.42}_{-0.48}$ &        SF &  $0.35^{+5.96}_{-0.15}$ & $0.004^{+0.04}_{-0.00}$ & $1.4^{+0.65}_{-1.18}$ &        2.59 \\
GRB070714B &                 $0.923$ &  $9.32^{+0.34}_{-0.31}$ &   $0.57^{+0.83}_{-0.84}$ &  $-8.75^{+1.13}_{-1.12}$ &        SF &  $1.12^{+5.19}_{-1.02}$ & $0.004^{+0.02}_{-0.00}$ & $0.9^{+0.80}_{-0.90}$ &        2.19 \\
 GRB070724 &                 $0.457$ & $10.01^{+0.28}_{-0.41}$ &   $0.59^{+1.07}_{-0.38}$ &  $-9.42^{+1.47}_{-0.55}$ &        SF &  $1.78^{+7.13}_{-1.67}$ &  $0.02^{+0.03}_{-0.02}$ & $0.3^{+1.13}_{-0.30}$ &        0.36 \\
 GRB070729 & $0.761^{+0.08}_{-0.08}$ &  $10.4^{+0.27}_{-0.13}$ &   $0.66^{+1.00}_{-1.11}$ &  $-9.74^{+0.93}_{-1.11}$ &        SF &  $0.35^{+3.54}_{-0.16}$ &  $0.05^{+0.00}_{-0.05}$ & $1.6^{+0.90}_{-1.36}$ &        1.20 \\
 GRB070809 &                 $0.473$ & $10.85^{+0.48}_{-0.12}$ &   $-0.1^{+0.96}_{-0.00}$ & $-10.95^{+0.93}_{-0.00}$ &         Q &  $1.12^{+7.79}_{-0.73}$ &  $0.02^{+0.03}_{-0.02}$ & $0.7^{+1.12}_{-0.70}$ &        4.69 \\
GRB070810B &                $0.0385$ & $10.73^{+0.03}_{-0.25}$ &  $-0.39^{+0.00}_{-0.00}$ & $-11.12^{+0.00}_{-0.00}$ &         Q & $12.59^{+0.00}_{-7.09}$ & $0.004^{+0.00}_{-0.00}$ & $0.4^{+0.21}_{-0.11}$ &        3.80 \\
 GRB071227 &                 $0.381$ & $10.53^{+0.31}_{-0.13}$ &    $0.8^{+0.91}_{-0.66}$ &  $-9.73^{+1.02}_{-0.88}$ &        SF &  $2.24^{+6.67}_{-1.77}$ &  $0.05^{+0.00}_{-0.04}$ & $1.2^{+1.25}_{-1.00}$ &        0.63 \\
 GRB080121 &                 $0.045$ &  $8.97^{+0.48}_{-0.26}$ &  $-0.58^{+0.79}_{-1.03}$ &  $-9.55^{+0.91}_{-0.77}$ &        SF & $1.41^{+11.18}_{-1.22}$ & $0.004^{+0.02}_{-0.00}$ & $0.2^{+0.81}_{-0.20}$ &        1.53 \\
 GRB080123 &                 $0.495$ & $10.03^{+0.15}_{-0.09}$ &   $0.39^{+0.48}_{-0.21}$ &  $-9.65^{+0.55}_{-0.21}$ &        SF &  $1.26^{+1.83}_{-0.77}$ &  $0.05^{+0.00}_{-0.03}$ & $0.2^{+0.67}_{-0.20}$ &        1.33 \\
 GRB090510 &                 $0.903$ &  $9.79^{+0.23}_{-0.19}$ &  $-0.13^{+0.94}_{-0.88}$ &  $-9.92^{+0.97}_{-0.88}$ &        SF &   $1.0^{+3.47}_{-0.79}$ & $0.004^{+0.05}_{-0.00}$ & $0.0^{+0.93}_{-0.00}$ &        0.78 \\
 GRB090515 &                 $0.403$ & $10.93^{+0.21}_{-0.40}$ & $-0.04^{+-0.17}_{-0.00}$ & $-10.97^{+0.09}_{-0.00}$ &         Q &  $5.01^{+3.90}_{-4.16}$ &  $0.02^{+0.03}_{-0.02}$ & $0.0^{+1.54}_{-0.00}$ &        0.30 \\
 GRB100117 &                 $0.914$ & $10.33^{+0.12}_{-0.20}$ &  $-0.47^{+0.20}_{-0.00}$ &  $-10.8^{+0.19}_{-0.00}$ &         Q &   $1.0^{+2.09}_{-0.53}$ & $0.008^{+0.04}_{-0.00}$ & $1.0^{+0.63}_{-0.69}$ &        1.06 \\
GRB100206A &                 $0.407$ & $10.79^{+0.28}_{-0.63}$ &   $0.68^{+1.93}_{-0.86}$ & $-10.11^{+2.49}_{-0.86}$ &        SF &  $3.98^{+4.93}_{-3.88}$ &  $0.05^{+0.00}_{-0.05}$ & $1.0^{+2.33}_{-0.84}$ &        0.14 \\
GRB100216A &                 $0.038$ &  $8.87^{+0.12}_{-0.20}$ &  $-0.89^{+0.41}_{-1.36}$ &  $-9.77^{+0.41}_{-1.35}$ &        SF &  $1.12^{+1.76}_{-0.85}$ & $0.004^{+0.01}_{-0.00}$ & $0.1^{+0.43}_{-0.10}$ &        1.08 \\
GRB100625A &                 $0.452$ &   $9.6^{+0.42}_{-0.18}$ &  $-0.69^{+1.26}_{-0.66}$ & $-10.29^{+1.28}_{-0.66}$ &        SF &  $1.41^{+7.50}_{-1.18}$ &  $0.05^{+0.00}_{-0.05}$ & $0.4^{+1.55}_{-0.40}$ &        1.15 \\
GRB100816A &                 $0.805$ & $10.37^{+0.43}_{-0.11}$ &    $1.0^{+0.97}_{-1.01}$ &  $-9.37^{+1.01}_{-1.19}$ &        SF &  $0.25^{+6.06}_{-0.12}$ & $0.008^{+0.04}_{-0.00}$ & $1.4^{+0.65}_{-1.40}$ &        4.57 \\
GRB101219A &                 $0.718$ &  $9.52^{+0.41}_{-0.43}$ &   $0.89^{+0.79}_{-0.82}$ &  $-8.63^{+1.02}_{-1.08}$ &        SF &  $0.79^{+6.29}_{-0.69}$ &  $0.05^{+0.00}_{-0.05}$ & $1.5^{+0.92}_{-1.15}$ &        1.70 \\
GRB101224A &                 $0.454$ &  $9.15^{+0.15}_{-0.24}$ &  $-0.01^{+0.61}_{-0.42}$ &  $-9.16^{+0.75}_{-0.30}$ &        SF &  $1.41^{+4.21}_{-1.27}$ & $0.008^{+0.04}_{-0.00}$ & $0.0^{+0.76}_{-0.00}$ &        3.41 \\
GRB110402A &                 $0.854$ &  $9.44^{+0.64}_{-0.59}$ &   $0.02^{+1.30}_{-0.73}$ &  $-9.42^{+1.68}_{-0.61}$ &        SF &  $1.78^{+4.53}_{-1.68}$ &  $0.02^{+0.03}_{-0.02}$ & $0.0^{+1.47}_{-0.00}$ &        0.86 \\
GRB111117A &                  $2.21$ &   $9.8^{+0.64}_{-0.31}$ &   $0.68^{+1.44}_{-0.25}$ &  $-9.13^{+1.51}_{-0.33}$ &        SF &  $0.22^{+2.60}_{-0.12}$ & $0.008^{+0.04}_{-0.00}$ & $0.1^{+1.01}_{-0.10}$ &        0.10 \\
GRB120305A &                 $0.225$ &  $9.28^{+0.16}_{-0.09}$ &  $-1.76^{+0.00}_{-0.00}$ & $-11.04^{+0.00}_{-0.00}$ &         Q &  $6.31^{+3.69}_{-1.94}$ & $0.004^{+0.00}_{-0.00}$ & $0.0^{+0.09}_{-0.00}$ &       20.20 \\
GRB120630A & $0.544^{+0.13}_{-0.08}$ &  $9.83^{+0.40}_{-0.40}$ &   $1.34^{+0.66}_{-0.81}$ &  $-8.49^{+0.88}_{-1.09}$ &        SF &  $0.18^{+8.73}_{-0.08}$ &  $0.05^{+0.00}_{-0.03}$ & $1.5^{+0.62}_{-1.10}$ &        0.28 \\
GRB120804A & $1.258^{+0.36}_{-0.37}$ & $10.46^{+0.26}_{-0.93}$ &  $-0.24^{+2.29}_{-0.00}$ &  $-10.7^{+3.08}_{-0.00}$ &         Q &  $4.47^{+1.84}_{-4.37}$ & $0.004^{+0.05}_{-0.00}$ & $0.9^{+2.94}_{-0.90}$ &        1.46 \\
GRB121226A & $2.103^{+0.20}_{-0.18}$ & $11.15^{+0.12}_{-0.47}$ &   $1.74^{+0.49}_{-0.43}$ &  $-9.42^{+0.94}_{-0.33}$ &        SF &  $3.16^{+0.00}_{-2.48}$ &  $0.05^{+0.00}_{-0.05}$ & $1.1^{+0.54}_{-0.44}$ &        5.26 \\
GRB130515A & $0.891^{+0.05}_{-0.05}$ &  $10.4^{+0.03}_{-0.08}$ &   $0.47^{+0.11}_{-0.88}$ &  $-9.93^{+0.09}_{-0.88}$ &        SF &  $0.56^{+0.07}_{-0.21}$ &  $0.02^{+0.02}_{-0.01}$ & $0.5^{+0.13}_{-0.27}$ &       29.70 \\
GRB130603B &                 $0.357$ &  $9.57^{+0.54}_{-0.11}$ &   $0.66^{+0.76}_{-0.67}$ &  $-8.91^{+0.84}_{-1.16}$ &        SF &   $0.2^{+8.71}_{-0.07}$ & $0.004^{+0.05}_{-0.00}$ & $1.4^{+0.33}_{-1.40}$ &        1.99 \\
GRB130822A &                 $0.154$ & $10.03^{+0.50}_{-0.44}$ &   $0.06^{+2.03}_{-1.10}$ &  $-9.97^{+2.35}_{-1.10}$ &        SF & $0.32^{+10.90}_{-0.22}$ & $0.008^{+0.04}_{-0.00}$ & $1.5^{+0.81}_{-1.50}$ &        0.38 \\
GRB140129B &                  $0.43$ &  $9.07^{+0.24}_{-0.67}$ &  $-0.58^{+1.53}_{-0.17}$ &  $-9.65^{+2.03}_{-0.32}$ &        SF &  $3.98^{+4.93}_{-3.88}$ &  $0.05^{+0.00}_{-0.05}$ & $0.0^{+1.93}_{-0.00}$ &        5.87 \\
GRB140622A &                 $0.959$ & $10.05^{+0.21}_{-0.09}$ &   $1.69^{+0.32}_{-0.35}$ &  $-8.35^{+0.37}_{-0.36}$ &        SF &  $0.16^{+1.04}_{-0.06}$ &  $0.05^{+0.00}_{-0.02}$ & $1.5^{+0.24}_{-0.32}$ &        1.22 \\
GRB140903A &                 $0.353$ & $10.26^{+0.55}_{-0.33}$ &   $1.44^{+0.90}_{-1.46}$ &  $-8.81^{+1.19}_{-1.80}$ &        SF &  $0.22^{+8.69}_{-0.12}$ & $0.008^{+0.04}_{-0.00}$ & $2.5^{+0.47}_{-2.38}$ &        0.89 \\
GRB140930B &                 $1.465$ &  $9.38^{+0.75}_{-0.10}$ &   $1.77^{+0.10}_{-1.29}$ &  $-7.61^{+0.00}_{-2.01}$ &        SF &   $0.1^{+4.37}_{-0.00}$ &  $0.02^{+0.03}_{-0.02}$ & $1.4^{+0.26}_{-1.40}$ &        0.32 \\
GRB141212A &                 $0.596$ &   $9.5^{+0.20}_{-0.15}$ &  $-0.61^{+1.11}_{-0.79}$ & $-10.11^{+1.02}_{-0.79}$ &        SF &  $0.89^{+3.00}_{-0.66}$ &  $0.05^{+0.00}_{-0.05}$ & $0.0^{+1.39}_{-0.00}$ &        1.11 \\
GRB150101B &                 $0.134$ & $10.85^{+0.50}_{-0.21}$ &  $-0.23^{+1.38}_{-0.00}$ & $-11.08^{+1.26}_{-0.00}$ &         Q &  $1.78^{+9.44}_{-1.07}$ & $0.008^{+0.04}_{-0.00}$ & $0.7^{+0.88}_{-0.70}$ &        1.40 \\
GRB150120A &                  $0.46$ & $10.08^{+0.34}_{-0.27}$ &   $0.51^{+0.84}_{-0.74}$ &  $-9.57^{+1.01}_{-1.03}$ &        SF &   $2.0^{+6.91}_{-1.82}$ & $0.008^{+0.03}_{-0.00}$ & $1.3^{+0.94}_{-0.95}$ &        3.67 \\
GRB150728A &                 $0.461$ &  $9.73^{+0.25}_{-0.19}$ &   $0.05^{+1.04}_{-0.38}$ &  $-9.68^{+1.10}_{-0.34}$ &        SF &  $0.89^{+3.09}_{-0.73}$ &  $0.05^{+0.00}_{-0.05}$ & $0.1^{+1.28}_{-0.10}$ &        0.72 \\
GRB150831A &                  $1.18$ &  $9.61^{+0.50}_{-0.30}$ &    $1.2^{+0.80}_{-2.31}$ &  $-8.41^{+0.80}_{-2.31}$ &        SF &  $0.14^{+4.87}_{-0.04}$ & $0.008^{+0.04}_{-0.00}$ & $1.9^{+0.77}_{-1.90}$ &        0.60 \\
GRB151229A & $1.879^{+0.80}_{-0.80}$ &  $9.85^{+1.12}_{-1.08}$ &   $2.25^{+0.75}_{-2.95}$ &  $-7.61^{+0.00}_{-2.94}$ &        SF &   $0.1^{+5.52}_{-0.00}$ &  $0.05^{+0.00}_{-0.05}$ & $2.2^{+0.47}_{-2.20}$ &        0.67 \\
GRB160303A & $1.095^{+0.28}_{-0.28}$ &  $9.51^{+0.37}_{-0.25}$ &  $-0.73^{+1.68}_{-0.50}$ & $-10.24^{+1.50}_{-0.50}$ &        SF &  $0.63^{+4.50}_{-0.45}$ &  $0.05^{+0.00}_{-0.05}$ & $0.5^{+1.55}_{-0.50}$ &        3.50 \\
GRB160408A & $1.483^{+0.72}_{-0.65}$ &  $8.92^{+1.94}_{-1.54}$ &  $-0.06^{+1.51}_{-1.66}$ &  $-8.98^{+1.37}_{-1.66}$ &        SF &  $2.82^{+3.49}_{-2.72}$ & $0.004^{+0.05}_{-0.00}$ & $0.0^{+1.43}_{-0.00}$ &        2.56 \\
GRB160411A & $0.732^{+0.37}_{-0.42}$ &  $8.62^{+0.65}_{-0.96}$ &   $0.95^{+0.22}_{-3.19}$ &  $-7.67^{+0.06}_{-3.19}$ &        SF &  $0.11^{+9.89}_{-0.01}$ & $0.004^{+0.05}_{-0.00}$ & $1.6^{+0.75}_{-1.60}$ &        4.53 \\
GRB160525B & $0.576^{+0.07}_{-0.08}$ &  $8.25^{+0.17}_{-0.51}$ &  $-0.08^{+0.42}_{-0.25}$ &  $-8.32^{+0.70}_{-0.26}$ &        SF &  $0.18^{+0.73}_{-0.08}$ &  $0.05^{+0.00}_{-0.05}$ & $0.0^{+0.31}_{-0.00}$ &       13.00 \\
GRB160624A &                 $0.484$ &  $9.69^{+0.48}_{-0.30}$ &   $1.19^{+0.62}_{-1.36}$ &   $-8.5^{+0.88}_{-1.70}$ &        SF &  $0.32^{+7.62}_{-0.22}$ & $0.008^{+0.04}_{-0.00}$ & $1.8^{+0.45}_{-1.80}$ &        0.94 \\
GRB160821B &                 $0.162$ &  $9.39^{+0.25}_{-0.58}$ &   $0.18^{+1.05}_{-1.86}$ &  $-9.21^{+1.59}_{-1.86}$ &        SF &  $4.47^{+6.75}_{-4.37}$ & $0.004^{+0.05}_{-0.00}$ & $0.2^{+0.93}_{-0.20}$ &        0.06 \\
GRB161001A & $0.776^{+0.07}_{-0.07}$ &  $10.1^{+0.22}_{-0.27}$ &  $-0.46^{+0.52}_{-0.28}$ & $-10.56^{+0.64}_{-0.28}$ &        SF &  $2.82^{+4.26}_{-2.10}$ & $0.004^{+0.02}_{-0.00}$ & $0.0^{+0.70}_{-0.00}$ &       18.25 \\
GRB161104A &                 $0.793$ & $10.38^{+0.32}_{-0.25}$ &  $-0.44^{+1.97}_{-0.02}$ & $-10.83^{+1.83}_{-0.01}$ &        SF &   $0.5^{+5.81}_{-0.29}$ & $0.004^{+0.05}_{-0.00}$ & $1.8^{+0.93}_{-1.80}$ &        0.59 \\
GRB170127B & $1.443^{+0.42}_{-0.33}$ &  $9.08^{+0.92}_{-0.93}$ &  $-0.12^{+0.87}_{-0.78}$ &  $-9.19^{+1.51}_{-0.82}$ &        SF &  $0.79^{+4.22}_{-0.69}$ & $0.004^{+0.05}_{-0.00}$ & $0.0^{+0.92}_{-0.00}$ &        4.53 \\
GRB170428A &                 $0.453$ &  $9.31^{+0.44}_{-0.30}$ &   $0.31^{+1.19}_{-0.41}$ &   $-9.0^{+1.38}_{-0.79}$ &        SF &  $0.25^{+8.66}_{-0.15}$ &  $0.05^{+0.00}_{-0.05}$ & $1.4^{+1.04}_{-1.01}$ &        4.85 \\
GRB170728A &                 $1.493$ & $10.98^{+0.17}_{-0.25}$ &    $0.6^{+1.01}_{-0.26}$ & $-10.38^{+1.03}_{-0.26}$ &        SF &  $3.16^{+0.82}_{-2.04}$ &  $0.05^{+0.00}_{-0.04}$ & $0.6^{+1.27}_{-0.60}$ &        6.74 \\
GRB170728B &                 $1.272$ &  $9.78^{+0.66}_{-0.25}$ &   $1.65^{+0.60}_{-1.86}$ &  $-8.14^{+0.53}_{-1.83}$ &        SF &   $0.2^{+4.81}_{-0.10}$ & $0.004^{+0.05}_{-0.00}$ & $1.1^{+0.71}_{-1.10}$ &        0.22 \\
 GRB170817 &               $0.00979$ & $10.26^{+0.44}_{-0.07}$ &  $-0.87^{+0.00}_{-0.00}$ & $-11.13^{+0.00}_{-0.00}$ &         Q &  $1.78^{+9.70}_{-0.87}$ &  $0.05^{+0.00}_{-0.04}$ & $0.4^{+0.90}_{-0.40}$ &        0.41 \\
GRB180418A & $1.094^{+0.78}_{-0.63}$ &  $9.13^{+1.61}_{-0.93}$ &   $1.51^{+1.27}_{-3.12}$ &  $-7.62^{+0.01}_{-3.12}$ &        SF &   $0.1^{+8.81}_{-0.00}$ & $0.004^{+0.05}_{-0.00}$ & $2.2^{+1.34}_{-2.20}$ &        7.56 \\
GRB180618A & $0.643^{+0.09}_{-0.09}$ &  $9.42^{+0.19}_{-0.00}$ &  $-1.47^{+0.00}_{-0.00}$ & $-10.89^{+0.00}_{-0.00}$ &         Q &  $0.56^{+0.18}_{-0.12}$ &  $0.05^{+0.00}_{-0.03}$ & $0.0^{+0.18}_{-0.00}$ &       24.33 \\
GRB180805B &                 $0.661$ &  $9.35^{+0.06}_{-0.03}$ &  $-0.62^{+0.39}_{-0.36}$ &  $-9.96^{+0.32}_{-0.37}$ &        SF &  $0.32^{+0.07}_{-0.06}$ &  $0.05^{+0.00}_{-0.03}$ & $0.0^{+0.34}_{-0.00}$ &        8.92 \\
GRB181123B &                  $1.75$ & $10.39^{+0.15}_{-0.78}$ &   $1.15^{+1.08}_{-1.02}$ &  $-9.24^{+1.63}_{-0.51}$ &        SF &  $3.16^{+0.39}_{-3.06}$ &  $0.05^{+0.00}_{-0.05}$ & $0.4^{+1.25}_{-0.40}$ &        1.16 \\
GRB191031D & $1.607^{+0.33}_{-0.24}$ & $10.78^{+0.23}_{-0.46}$ &   $0.67^{+1.78}_{-0.53}$ & $-10.11^{+2.12}_{-0.46}$ &        SF &  $3.98^{+0.49}_{-3.87}$ &  $0.05^{+0.00}_{-0.05}$ & $0.4^{+1.99}_{-0.40}$ &        2.08 \\
GRB200219A & $0.457^{+0.18}_{-0.21}$ & $10.77^{+0.36}_{-0.99}$ &   $0.35^{+1.77}_{-0.53}$ & $-10.42^{+2.25}_{-0.53}$ &        SF &  $5.62^{+4.38}_{-5.49}$ &  $0.05^{+0.00}_{-0.05}$ & $0.2^{+2.85}_{-0.20}$ &       27.86 \\
GRB200411A & $1.145^{+0.07}_{-0.06}$ & $10.57^{+0.21}_{-0.13}$ &   $0.89^{+0.71}_{-0.24}$ &  $-9.68^{+0.58}_{-0.19}$ &        SF &  $0.89^{+1.06}_{-0.65}$ &  $0.05^{+0.00}_{-0.05}$ & $0.1^{+0.90}_{-0.10}$ &        3.90 \\
GRB200522A &                 $0.554$ &  $9.73^{+0.19}_{-0.24}$ &   $0.53^{+0.51}_{-0.67}$ &   $-9.2^{+0.65}_{-0.54}$ &        SF &  $0.79^{+7.15}_{-0.63}$ &  $0.02^{+0.03}_{-0.02}$ & $0.2^{+0.70}_{-0.20}$ &        0.38 \\
GRB201221D &                 $1.055$ &  $9.79^{+0.14}_{-0.63}$ &   $0.59^{+1.14}_{-0.10}$ &  $-9.19^{+1.58}_{-0.25}$ &        SF &  $4.47^{+1.15}_{-4.37}$ &  $0.05^{+0.00}_{-0.05}$ & $0.0^{+1.42}_{-0.00}$ &        5.09 \\
GRB210323A &                 $0.733$ &  $8.79^{+0.54}_{-0.86}$ &   $-0.3^{+0.84}_{-0.29}$ &  $-9.09^{+1.47}_{-0.60}$ &        SF &  $3.55^{+3.53}_{-3.45}$ &  $0.05^{+0.00}_{-0.05}$ & $0.0^{+0.93}_{-0.00}$ &        5.29 \\
GRB210510A &                 $0.221$ &  $9.43^{+0.31}_{-0.34}$ &   $1.29^{+0.42}_{-0.51}$ &  $-8.14^{+0.53}_{-0.67}$ &        SF &  $0.11^{+1.89}_{-0.01}$ &  $0.05^{+0.00}_{-0.04}$ & $2.4^{+0.43}_{-0.63}$ &        1.73 \\
GRB210726A & $0.458^{+0.17}_{-0.18}$ &  $9.11^{+0.50}_{-0.75}$ &   $0.12^{+1.33}_{-1.96}$ &  $-8.99^{+1.37}_{-1.96}$ &        SF &  $2.82^{+6.73}_{-2.72}$ &  $0.02^{+0.03}_{-0.02}$ & $0.0^{+1.28}_{-0.00}$ &        2.32 \\
GRB210919A &                 $0.242$ &  $10.4^{+0.12}_{-0.49}$ &  $-0.64^{+0.00}_{-0.00}$ & $-11.04^{+0.00}_{-0.00}$ &         Q &  $7.94^{+2.06}_{-7.11}$ &  $0.02^{+0.03}_{-0.02}$ & $0.0^{+1.36}_{-0.00}$ &        8.66 \\
GRB211023B &                 $0.862$ &  $9.09^{+0.99}_{-0.34}$ &  $-0.19^{+1.98}_{-1.09}$ &  $-9.28^{+1.67}_{-1.04}$ &        SF &   $0.5^{+5.81}_{-0.40}$ &  $0.02^{+0.03}_{-0.02}$ & $0.0^{+2.20}_{-0.00}$ &        0.54 \\
GRB211211A &                $0.0763$ &  $8.84^{+0.17}_{-0.43}$ &  $-1.03^{+0.81}_{-0.80}$ &  $-9.87^{+1.13}_{-0.60}$ &        SF &  $5.62^{+6.97}_{-5.40}$ & $0.004^{+0.02}_{-0.00}$ & $0.1^{+0.88}_{-0.10}$ &        0.46 \\
GRB211227A &                 $0.228$ & $10.23^{+0.13}_{-0.10}$ &   $-0.5^{+1.36}_{-0.31}$ & $-10.73^{+1.27}_{-0.31}$ &        SF &   $0.4^{+0.89}_{-0.15}$ &  $0.05^{+0.00}_{-0.05}$ & $1.3^{+0.70}_{-0.60}$ &        4.66 \\
GRB221120A &  $0.08^{+0.04}_{-0.03}$ & $10.06^{+0.46}_{-0.75}$ &   $0.92^{+1.10}_{-0.92}$ &  $-9.14^{+1.53}_{-0.73}$ &        SF &  $3.98^{+8.61}_{-3.88}$ &  $0.05^{+0.00}_{-0.02}$ & $1.2^{+0.96}_{-0.62}$ &        5.65 \\

\midrule
\multicolumn{10}{l}{\textbf{Notes:}} \\
\multicolumn{10}{p{6in}}{$^{a}$ Galaxy types according to specific star formation rate. `Q' denotes quiescent galaxies, while `SF' denotes star-forming galaxies.} \\
\end{longtable*}


\subsection{Control Sample Galaxies Definition} \label{sec:comparison}

\subsubsection{Galaxies in the COSMOS Field} \label{sec:cosmos_galaxies}

To quantitatively assess the characteristics of short and hybrid GRB host galaxies, we set a control sample of non-host galaxies. Here, we utilize photometric and spectroscopic catalogs for galaxies in the COSMOS field. The COSMOS field is suitable for establishing a well-defined control group as it offers photometric information with a wide wavelength range from the UV to MIR, as well as some spectral analysis data.

COSMOS2020 is one of the deepest photometric catalogs available, with an $i$-band 3$\sigma$ sensitivity limit of 27.0 mag across an effective survey area of 1.27 deg$^2$ \citep{2022ApJS..258...11W}. The COSMOS2020 data cover a moderately wide area and contain accurate photometric redshifts ($|\Delta z|<0.025(1+z)$ at $i<25\mathrm{AB}$), making it suitable for constructing the control sample.

The catalog contains approximately one million sources. To minimize contamination in photometry values (e.g., bright stars affecting galaxy photometry), we selected those with \texttt{FCOMBINED} values of 0 \citep{2023A.A...677A.184W}. Furthermore, we chose sources with \texttt{ACSmuClass} values of 1 to ensure a collection of purely extended sources. These constraints return a refined sample of 524,901 galaxies.

The catalog offers information such as photometric redshifts, stellar masses, SFRs, and rest-frame absolute magnitudes from SED fitting.  The method of estimating galaxy properties in the existing COSMOS2020 catalog, however, differs from that used for our sGRB host galaxies. Therefore, we performed a new SED fitting for the refined sample of COSMOS2020 galaxies following the same method we used for the sGRB host galaxy sample.

The COSMOS2020 photometric data are collected from various telescopes, including $u$-band from Canada-France-Hawaii Telescope (CFHT), $grizy$ from Subaru~/~HyperSuprime-Cam (HSC), $YJHK_{s}$ from UltraVISTA, $FUV$ and $NUV$ from GALEX, and $F814W$ from HST/ACS, which were listed in the {\fontfamily{pcr}\selectfont THE FARMER} version catalog. Additionally, we incorporated spectroscopic redshifts provided by the zCOSMOS \citep{2007ApJS..172...70L,2009ApJS..184..218L} and hCOSMOS \citep{2018ApJS..234...21D} for 8,673 galaxies matched with the photometric catalog within 1 arcsec. For duplicate targets, we referred the hCOSMOS which has higher accuracy.

We fixed the collected photometric and spectroscopic redshift and conducted SED fitting using \texttt{FAST}. The overall processes were similar to the sGRB host analyses in Section \ref{sec:sedmodelling}, but without Monte-Carlo repetitions. When we exclude about 0.7 \% of poorly fitted samples with $\chi^2_{reduced} > 5$, the redshift and stellar mass generally agreed well with the values from the COSMOS2020 catalog.

We estimated the rest-frame absolute magnitudes of the fitted galaxies as given below. We first corrected for the effects of redshift on the wavelength and flux of the best-fit model and then performed synthetic photometry using the SDSS $r$-band filter curve. This calculation method includes the concept of the conventional K-correction.

\begin{figure}
    \centering
    \includegraphics[width=\columnwidth]{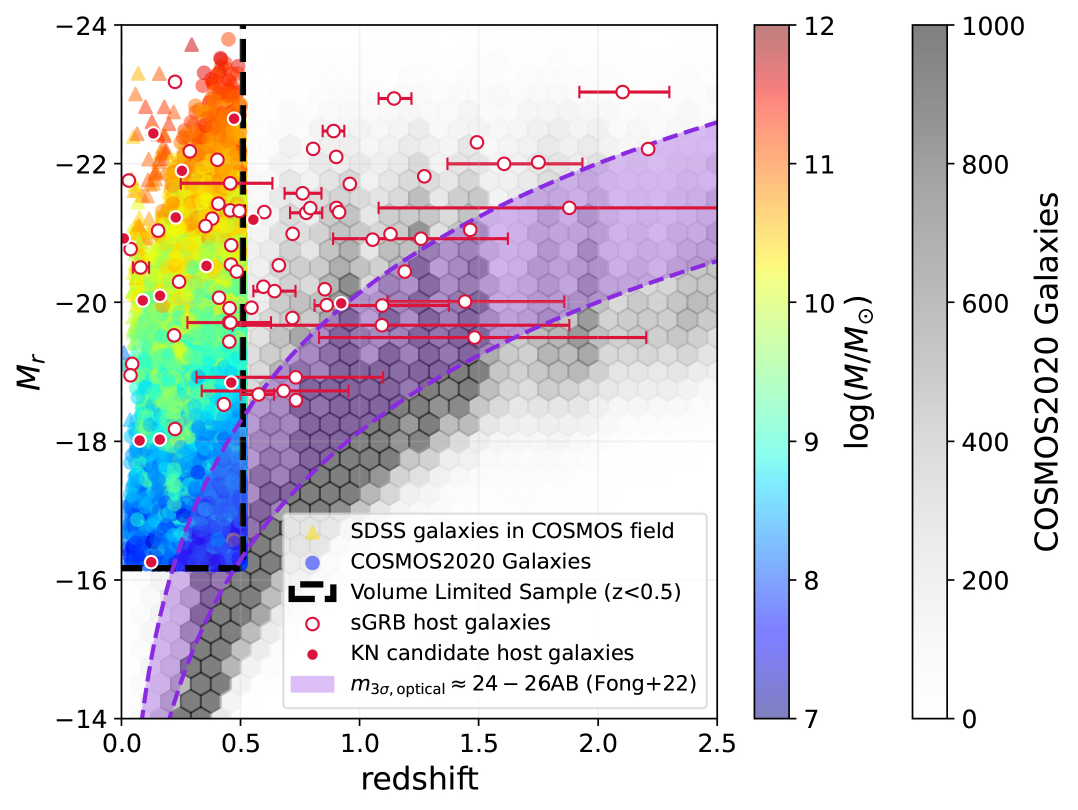}
    \caption{Redshift versus rest-frame absolute r-magnitude distribution of COSMOS field galaxies. The dotted-line box represents the volume-limited sample based on $z<0.5$ and  $M_{r}<-16.22 \ \mathrm{mag}$. The magnitudes of galaxies represented by colored markers were calculated according to the method described in Section~\ref{sec:sedmodelling}.  The gray hexagon bins indicate the \texttt{lpzBEST} and \texttt{lpRMAG} parameters in the COSMOS2020 {\fontfamily{pcr}\selectfont THE FARMER} catalog}. The purple-shaded region represents the typical sGRB host survey limiting magnitude \citep{2022ApJ...940...56F}.
    \label{fig:vlset}
\end{figure}

While the hCOSMOS catalog is valuable for providing redshift information for faint galaxies, it lacks completeness for bright galaxies. SDSS DR12 is useful because it offers spectroscopic redshifts for bright galaxies \citep{2015ApJS..219...12A}. We selected SDSS galaxies residing in the COSMOS field and collected their $ugriz$ photometric data and spectroscopic redshifts when available. We identified 151 galaxies occupying $r<17.7$ mag and included their spectroscopic redshifts to increase the spectroscopic redshift completeness at the bright end of the COSMOS2020 sample. The absolute magnitude distributions of COSMOS2020 and SDSS DR12 galaxies can be seen in Figure \ref{fig:vlset}.

\subsubsection{Volume-limited Sample} \label{sec:vl_sample}

Figure \ref{fig:vlset} shows the absolute magnitudes versus redshifts of host galaxies. At low redshifts ($z \lesssim 0.5$),  the majority of host galaxies lie well above the detection limit of the follow-up imaging observations to find GRB host galaxies, but at higher redshifts, the host galaxy magnitudes cut off near the detection limit. This suggests that the higher redshift ($z \gtrsim 0.5$) host galaxy sample could be incomplete due to the imaging depths. Furthermore, the classification of GRBs and assessing their association with host galaxies become more ambiguous at higher redshifts \citet{2013ApJ...764..179B}. In the near future, the NS-involved GW events are expected to be observed at low redshift, where the properties of their host galaxies would be of interest \citep{2018LRR....21....3A}. 

\begin{figure*}
    \centering
    \includegraphics[width=\linewidth]{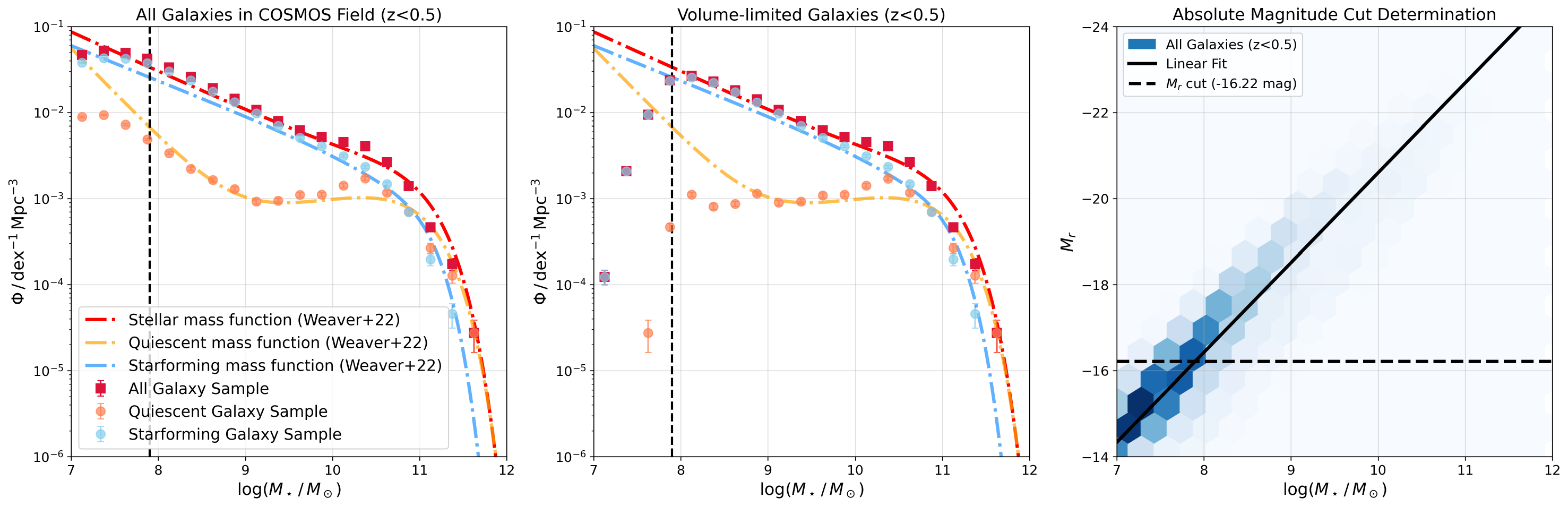}
    \caption{The galaxy stellar mass function for the COSMOS field galaxies at $z<0.5$, with errors represented solely by Poisson noise (\textbf{Left}). The same figure but with a volume-limited COSMOS sample defined by an absolute magnitude cut-off at $M_{r}<-16.22 \ \mathrm{mag}$ (\textbf{Middle}). The relationship between stellar mass and $r$-band absolute magnitude, with a solid line depicting the linear fit for all galaxies at $z<0.5$. The dotted line represents the absolute magnitude cut corresponding to a stellar mass of $\log(M_{*}/M_{\odot}) = 7.9$ (\textbf{Right}). In the first two panels, the orange and sky-blue data points refer to the passive galaxies and star-forming galaxies, respectively, which are classified by Equation~\ref{eqn:passive}.} 
    \label{fig:GSMF}
\end{figure*}

Therefore, we decided to construct a volume-limited sample for $z<0.5$ for the analysis of the host galaxy properties. A volume-limited sample refers to a group that is not biased towards nearby and brighter galaxies by giving an absolute magnitude cut within a given redshift range. Although there could be a certain amount of evolution in galaxies between $z=0-0.5$, limiting the sample to a lower redshift (i.e., smaller volume), for example, $z<0.3$, would leave only 22 host galaxies available, and hence we opted to set the redshift limit at 0.5 where 39 host galaxies will be available for the analysis.

To define the volume-limited sample, we must set an absolute magnitude cut in the $z-M_{r}$ plane (Figure \ref{fig:vlset}). This magnitude cut determines the stellar mass completeness, and to determine the cut, we plotted the galaxy stellar mass function (GSMF). In the left panel of  Figure \ref{fig:GSMF}, the number density of the $z<0.5$ COSMOS2020 sample as a function of a mass bin is also presented. By comparing this with the result from \citet{2023A.A...677A.184W}, we could confirm that the data set is not fully complete down to $10^7 M_{\odot}$.  Therefore, we chose the lower absolute magnitude limit corresponding to $10^{7.9} M_{\odot}$ to cover the minimum stellar mass of sGRB host galaxy while maintaining the completeness of the data. In the right panel of Figure \ref{fig:GSMF}, we demonstrate that selecting COSMOS field galaxies with  $M_r < -16.22 \ \mathrm{mag}$ results in a volume-limited COSMOS sample of 35,122 galaxies that adheres to the GSMF down to $\log(M_{\ast}/M_{\odot}) \simeq 8$. This selection criterion excludes about 1.3\% of galaxies within the stellar mass range of $8 < \log(M_{\ast}/M_{\odot}) < 9$. 

 Applying such an absolute magnitude cut to our host galaxies ensures that the entire sample of 39 host galaxies with $z<0.5$ is included. This allows us to compare the short and hybrid GRB host galaxies and the control sample under the same density conditions. If we had set a brighter magnitude cut that does not include the mass of faint galaxies like the host galaxy of GRB 060614, we would have compared a smaller number of samples. However, the impact of a single object on the overall results is not significant. The differences in results due to the absolute magnitude cut will be discussed later.

\subsubsection{Passive Galaxy Definition} \label{sec:passive_cut}

During the SED fitting process, we calculated each galaxy's star formation rate (SFR). However, we note that some galaxies have log(SFR) values that tend towards negative infinity, indicating minimal star formation activity. In reality, it is challenging to discern differences between galaxies with low SFRs using broad-band SED due to their negligible variations. Therefore, taking these small values at face value would be unphysical. Moreover, setting SFR to zero complicates the derivation of an analytical solution. To address this issue, we introduce a lower limit and propose using Equation \ref{eqn:passive} as a suitable criterion.

\begin{equation}\label{eqn:passive}
\mathrm{sSFR} < \frac{1}{10\times t(z)}
\end{equation}

In the above equation, $t(z)$ represents the universe age at redshift z. Setting such a lower limit of specific SFR (SFR per unit stellar mass) implies that it would take more than 10 times the age of the universe to build the mass of that galaxy assuming current SFR. This makes it a suitable criterion to evaluate galaxies where the SFR has been quenched. While COSMOS2020 used color cut to define passive galaxies sample, both criteria return similar passive galaxies of $\log(\mathrm{sSFR/yr})<10^{-11}$. On the other hand, some other studies try to define passive galaxies as ones with $\mathrm{sSFR}<1/[3\times t(z)]$ or higher values. Using such criteria assigns higher sSFR values to passive galaxies, which would influence the assessment of sSFR contribution to the BNS merger rate. We will address this effect in the discussion section.

\section{Result} \label{sec:result}

In this section, we present the properties of host galaxies estimated from the SED fitting processes in the previous section. From the distributions of stellar masses and star formation activities of the host galaxies, we can determine how they deviate from those of general galaxies. By constructing a weight function that compensates for the deviation, we would be able to evaluate the contributions of host galaxy properties that reflect the effects of binary neutron star merger rate.

\subsection{The Stellar Mass Distribution of  Short and Hybrid GRB Host Galaxies} \label{sec:properties}

\begin{figure}
\begin{center}
\includegraphics[width=\columnwidth]{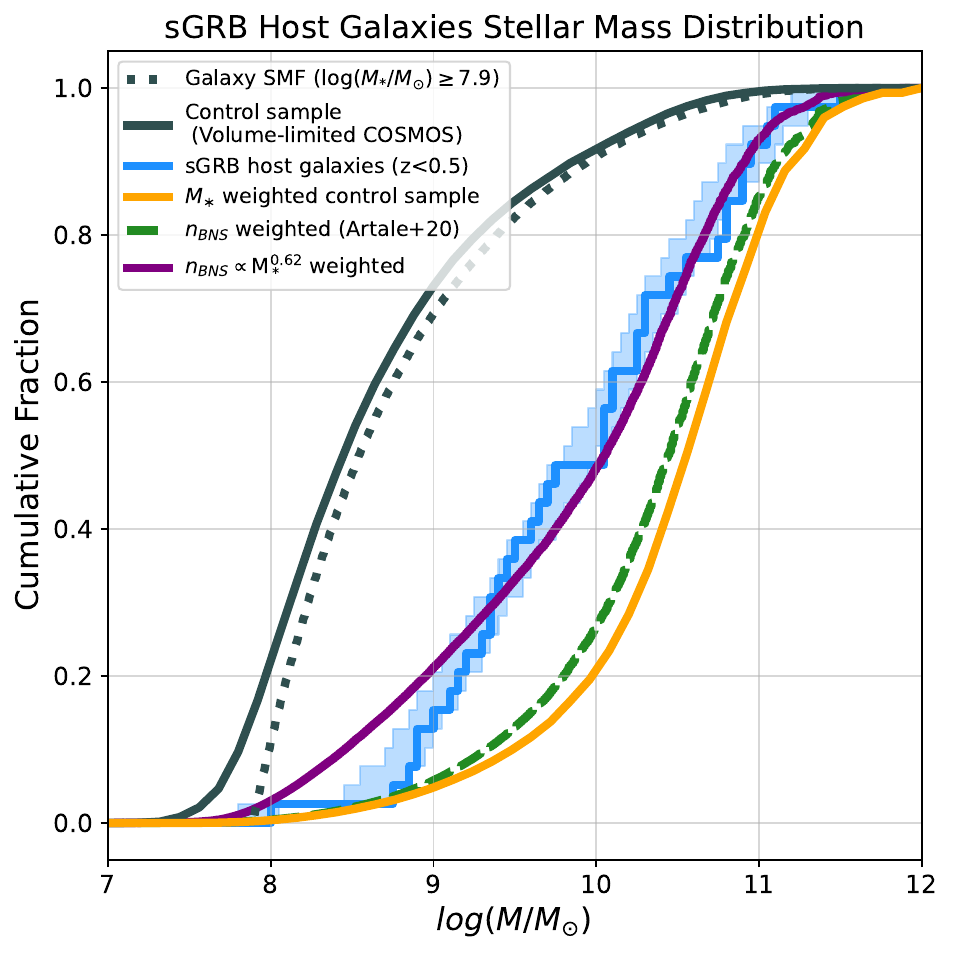}
\caption{The cumulative distributions of stellar mass for short and hybrid GRB host galaxies compared with a volume-limited sample of galaxies in COSMOS field, applying various weighting schemes. The dotted gray line indicates the theoretical cumulative distribution of galaxies adhering to the stellar mass function for redshifts less than 0.5, with a stellar mass lower limit of $10^{7.9} M_{\odot}$ \citep{2023A.A...677A.184W}. The continuous gray line tracks the cumulative distribution of a control sample galaxies. The blue stepped line represents the cumulative distribution for 39 GRB host galaxies with redshifts below 0.5, while the blue shaded region indicates the uncertainty range in stellar mass estimates. The orange line shows the cumulative distribution of the control sample, with each galaxy's contribution weighted by its stellar mass. The green long-dashed line illustrates the predicted host mass distribution based on the local BNS merger rate suggested by \citet{2020MNRAS.491.3419A}. The purple line represents the control sample weighted by $M_{\ast}^{0.62}$, aimed at replicating the distribution pattern of the host galaxies.
\label{fig:cummulative0}}
\end{center}
\end{figure}

Stellar mass is one of the major parameters expected to closely relate to the BNS merger rate. The stellar mass is the outcome of a galaxy's star formation history (SFH), which influences the formation rate of the progenitor system and the delay in its merging. In this context, comparing and analyzing the distributions of sGRB host galaxies and general field galaxies regarding their stellar mass is essential. By doing so, we can evaluate the differences as the contribution of the BNS merger rate.

If the binary neutron star (BNS) merger rate is directly proportional to the stellar mass of the host galaxy, as suggested by some theoretical studies \citep{2018MNRAS.481.5324M}, then the mass distribution of the host galaxies would follow a mass-weighted distribution. Under such a model, a $10^{10} M_{\odot}$ galaxy would exhibit a relative likelihood a hundred times greater of being identified as a sGRB host than a $10^{8} M_{\odot}$ galaxy. Consequently, this would skew the frequency in the cumulative distribution towards more massive galaxies, deviating from the intrinsic distribution that typically follows the galaxy stellar mass function.

Figure \ref{fig:cummulative0} shows the following results. Compared to the volume-limited COSMOS field galaxies, which almost follow galaxy stellar mass function with the lower limit of $10^{7.9} M_{\odot}$, short and hybrid GRB host galaxies obviously lean toward a massive part of the distribution.  However, they are not as massive as the mass-weighted distribution of the control sample, where the contribution of each galaxy is weighted by its stellar mass. This implies that the expected BNS merger rate in each galaxy is not directly proportional to its stellar mass.

Instead, if a weight of $M_{\ast}^{0.62}$ is applied to the control sample, it then appears similar to the observed mass distribution of the host galaxies. This is the result of finding the highest p-value returned when the exponent of the mass in the weight function was varied in 0.01 increments from 0 to 1 when the Kolmogorov-Smirnov (K-S) test was conducted with the host galaxies. As seen in these results, the environment of short and hybrid GRB occurrence has a weaker correlation with host stellar mass than a prediction of a linear relationship would suggest. Consequently, this implies the need to pay attention to low mass galaxies in BNS merger follow-up observations.

\subsection{BNS Merger Rate Proxy} \label{sec:merger_rate}

Given that host galaxy properties can potentially affect the BNS merger rate, it is plausible to consider an alternative form of the weight function. \citet{2020MNRAS.491.3419A} utilized hydrodynamic simulations and synthetic binary models to investigate the properties and cosmic evolution of BNS merger host galaxies. Their results suggest that host galaxies at $z>1$ have active star formation activities, evolving through galaxy mergers during the long delay times of BNSs. In the local universe ($z=0.1$), they predict that BNS merger host galaxies should be quiescent and massive. They propose a predictive formula for the local BNS merger rates as $\log(n_{\mathrm{BNS}}/\mathrm{Gyr}) = 0.800 \times \log(M_{\ast}/M_{\odot}) + 0.323 \times \log(\mathrm{SFR/M_{\odot} yr^{-1}}) - 3.555$, emphasizing the importance of stellar mass for predicting BNS merger rates for a galaxy. To directly compare with their prediction and the observed properties of short and hybrid GRB host galaxies, we include the star formation rate as a parameter in our weight function.

\begin{figure}
\begin{center}
\includegraphics[width=\columnwidth]{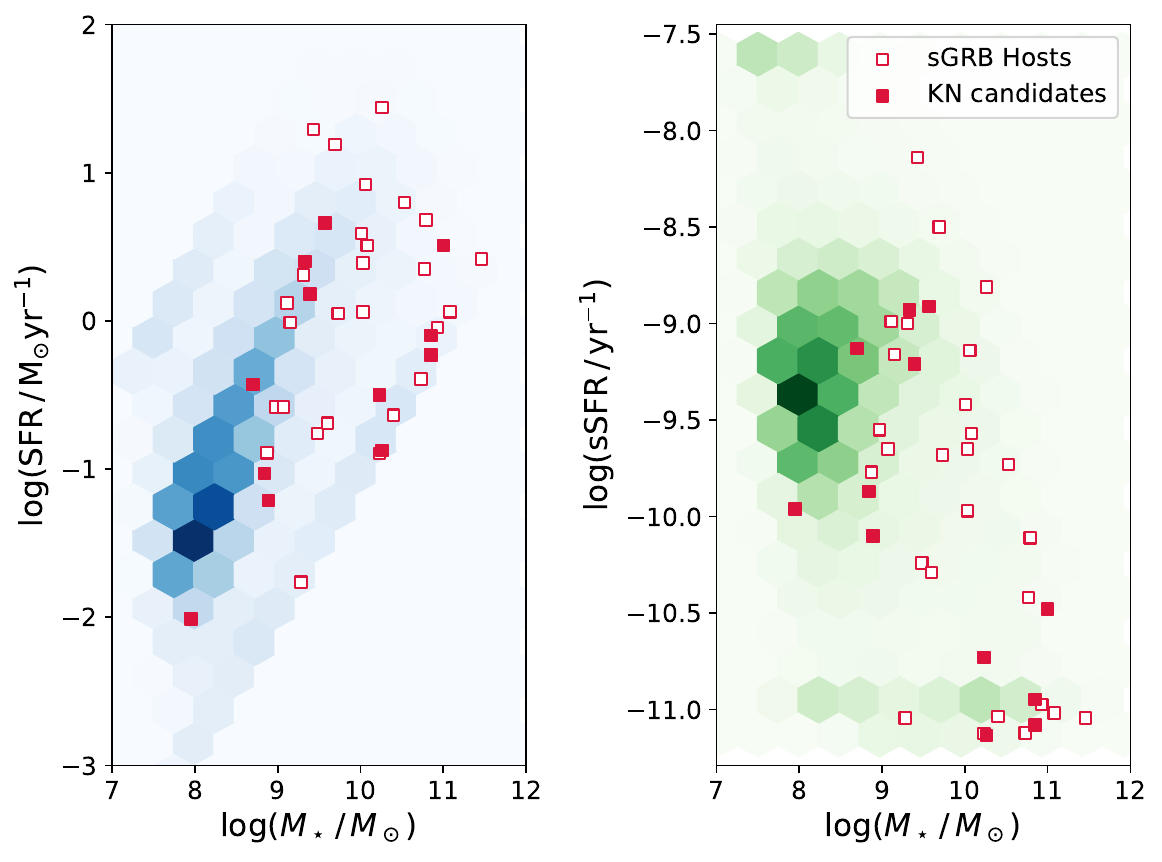}
\caption{The host galaxies and field galaxies distribution in the $M_{\ast}$-SFR plane (\textbf{Left}) and $M_{\ast}$-sSFR plane (\textbf{Right}). In both diagrams, the hexagon bins in the background represent the distribution of volume-limited sample of galaxies in the COSMOS field. The lower limit of sSFR is set by Equation \ref{eqn:passive}.
\label{fig:SFRvssSFR}}
\end{center}
\end{figure}

One thing to consider is that those parameters are not mutually independent; for example, the relationship between stellar mass and SFR for star-forming galaxies is well established (e.g., \citealt{2008MNRAS.385..147D,2010MNRAS.408.2115M}), a correlation also apparent in the left panel of Figure \ref{fig:SFRvssSFR}. Thus, to minimize the interdependencies between variables and clearly outline each term's contribution, we have chosen to employ the specific SFR as the second parameter. Therefore, we aim to use Equation \ref{eqn:gwrate} as the predictive formula for the BNS merger rate and apply it as a weight function for the mass distribution.

\begin{equation} \label{eqn:gwrate}
\begin{aligned}
\log(n_{\mathrm{BNS}}/\mathrm{Gyr}) = a \times \log(M_{\ast}) + b \times \log(\mathrm{sSFR/yr}) + C
\end{aligned}
\end{equation}

\begin{figure}
\begin{center}
\includegraphics[width=\columnwidth]{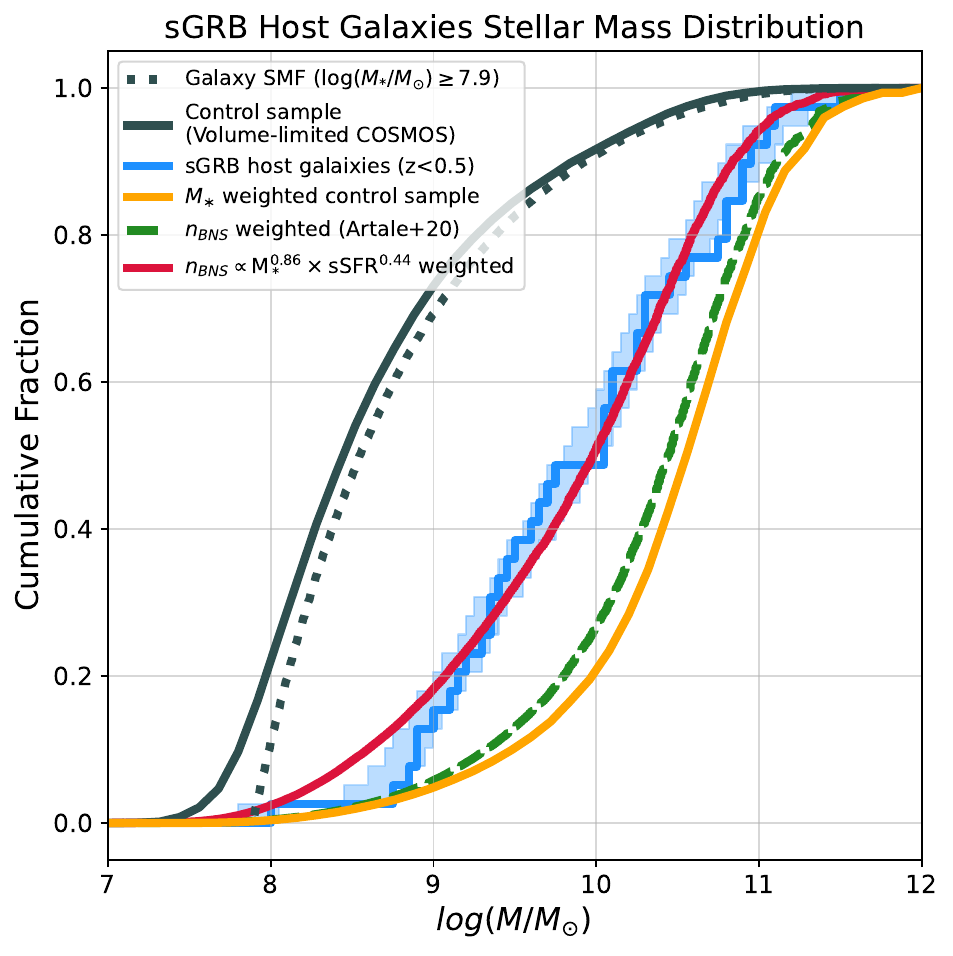}
\caption{This figure presents the cumulative stellar mass distribution of short and hybrid GRB host galaxies akin to Figure \ref{fig:cummulative0}. It highlights the weighted distribution (red line) that accounts for both stellar mass and sSFR (Equation \ref{eqn:gwrate}), which optimally reproduces the observed distribution of short and hybrid GRB host galaxy stellar masses.
\label{fig:cummulative}}
\end{center}
\end{figure}

Assuming a certain set of coefficients $(a, b)$ of this equation, we can weight the number count of each galaxy in control sample with its stellar mass and sSFR. The resulting weighted distribution can be compared to short and hybrid GRB host galaxy distribution using Kolmogorov-Smirnov (K-S) test (Figure \ref{fig:cummulative}).

\begin{figure}
\begin{center}
\includegraphics[width=\columnwidth]{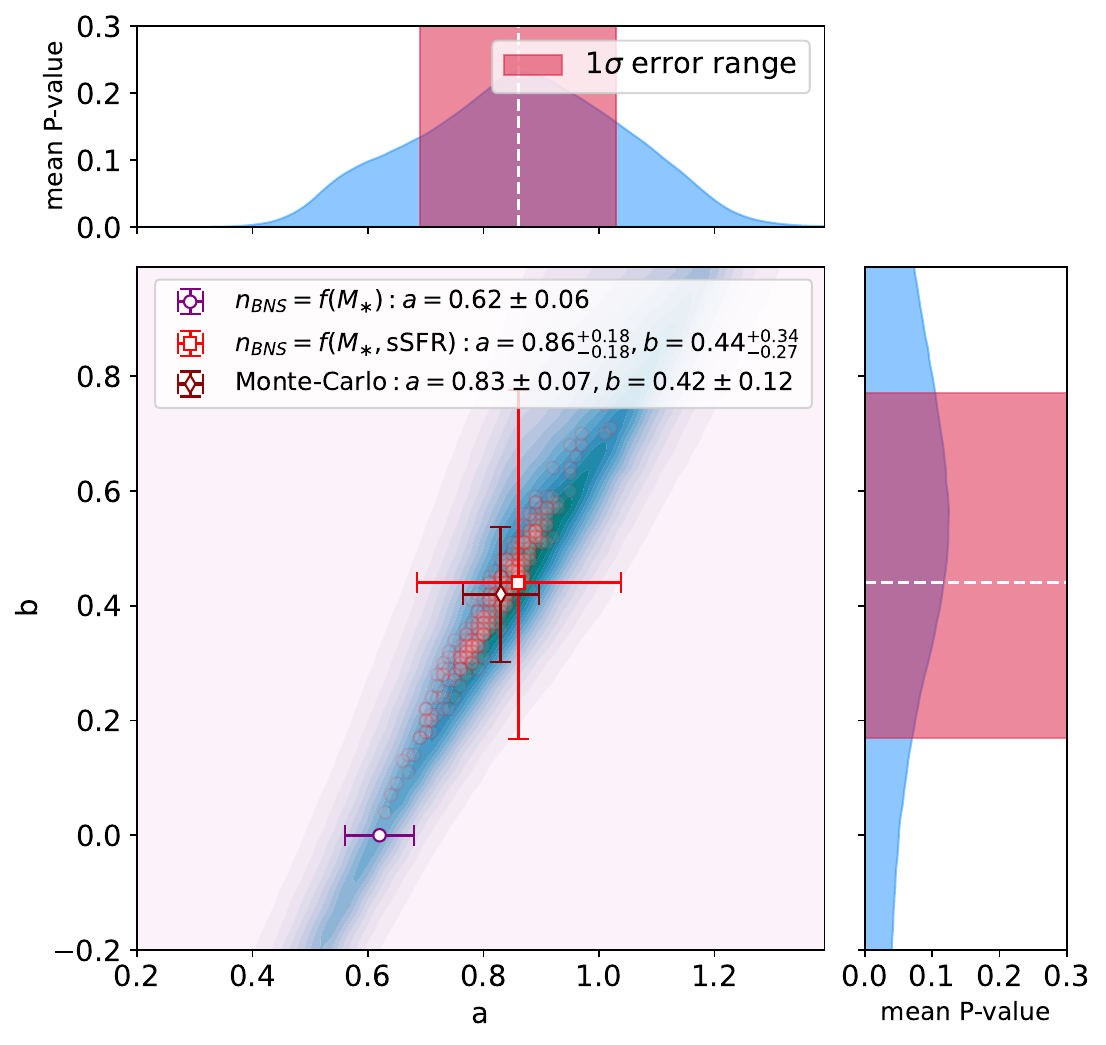}
\caption{This figure displays the p-value distribution resulting from the K-S test comparing 39 short and hybrid GRB host galaxies to a volume-limited COSMOS galaxies at $z<0.5$. The distribution is Gaussian smoothed at the $1\sigma$ level to explicate the trend. The purple circle marks the maximum p-value when the analysis is restricted solely to the stellar mass as the variable, while the red square represents the maximum p-value achieved when both stellar mass and sSFR are incorporated into the weighting function. The error bar for the red square delineate the standard deviations of the marginalized distributions along each axis. The semi-transparent red dots represent best-fit values of Monte-Carlo simulations, where each simulation accounts for the stellar mass errors of short and hybrid GRB host galaxies. The brown diamond indicates the median value of these best-fit values from the repetitions.
\label{fig:grid_search}}
\end{center}
\end{figure}

To find the optimal set of $(a, b)$, we can conduct a grid-searching method. Specifically, we perform the K-S test iteratively by varying the value of $a$ from 0.2 to 1.4 and $b$ from -0.2 to 1.0, with the step size of 0.01. The chosen parameter space encompasses the mass-prioritized prediction and includes generous variations to account for the contribution of the star formation rate. We confirmed that this range adequately covers the region around the maximum points and that there are no other local maxima beyond these bounds. This approach allows for a thorough investigation of the p-value for each pair of $(a, b)$. The result, shown in Figure \ref{fig:grid_search}, indicates the highest p-value at $(a, b)$ = ($0.86^{+0.18}_{-0.18}, 0.44^{+0.34}_{-0.27}$). To estimate the error of the coefficients, we marginalized the p-value distribution to the $a$ and $b$ axes and measured the 1$\sigma$ standard deviation. This suggests that $0.86 \times \log(M_{\ast}/M_{\odot}) + 0.44 \times \log(\mathrm{sSFR/yr})$ can be a proxy for the BNS merger rate that reproduces the observed characteristics of short and hybrid GRB host galaxies.

 To account for the effect of uncertainty in the stellar mass estimated from SED fitting, we performed random sampling 1,000 times using a skew-normal distribution estimated from the median and 1$\sigma$ upper and lower limits represented in Table \ref{tab:fitresult}. The stellar mass distribution accounting for uncertainty is shown as the shaded region in Figure \ref{fig:cummulative}, representing the range from the 10th to the 90th percentile. After 1,000 Monte Carlo simulations, the median and standard deviation of the optimally estimated $(a, b)$ were found to be $(0.83 \pm 0.07, 0.42 \pm 0.12)$. This result is well within the error range of the optimal coefficients estimated without considering the uncertainty.

\begin{figure*}
\begin{center}
\includegraphics[width=\textwidth]{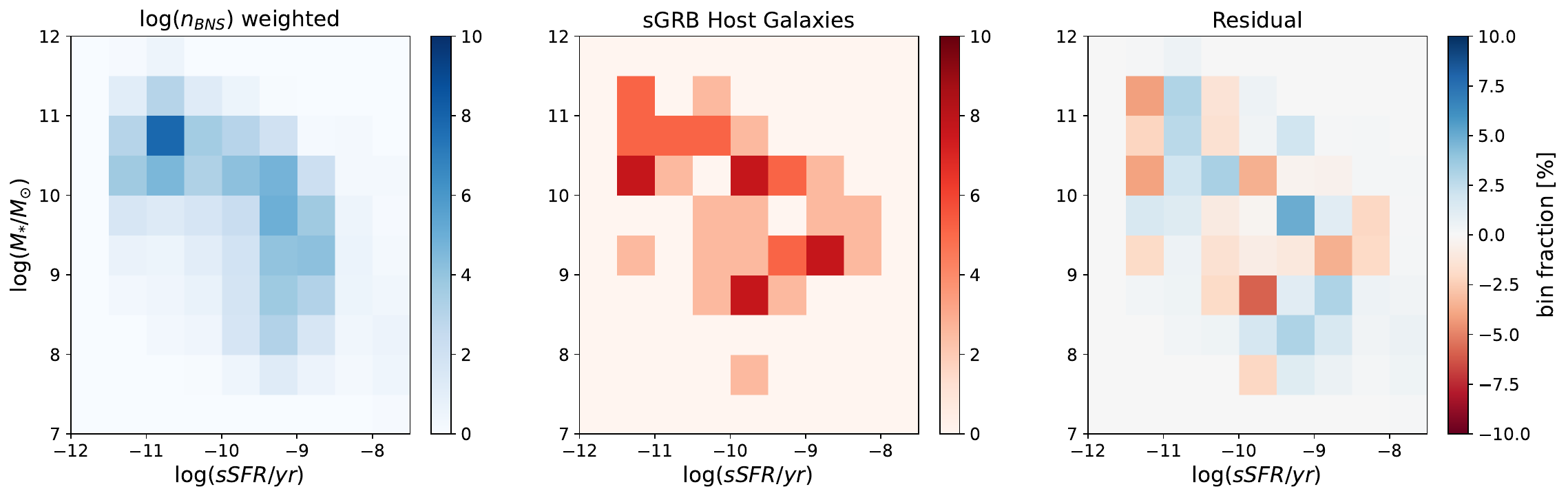}
\caption{The normalized distributions of control sample and short and hybrid GRB host galaxies within the $\log(\mathrm{sSFR/yr})-\log(M_{\ast}/M_{\odot})$ plane. The left panel illustrates the number density of volume-limited COSMOS sample, adjusted by the weight function of  $\log(n_{\mathrm{BNS}}/\mathrm{Gyr})=0.67\times\log(M_{\ast}/M_{\odot})+0.14\times\log(\mathrm{sSFR/yr})+C$, which is determined to minimize the sum of squared residuals. The central panel depicts the distribution of short and hybrid GRB host galaxies, and the right panel shows the residual distribution after the optimization. The color scales have been normalized to represent the fraction of counts in each bin relative to the total sample size, allowing bin fractions to be interpreted as percentages.
\label{fig:2dim}}
\end{center}
\end{figure*}

\subsection{Short and Hybrid GRB Host Galaxy Distribution on a Mass-sSFR Plane} \label{2d fitting}

In the previous section, we performed K-S tests on the one-dimensional stellar mass distribution. Since we consider the stellar mass and specific star formation rate (sSFR) as primary indicators for predicting the BNS merger rate, a more robust approach involves comparing host galaxies with the volume-limited COSMOS galaxies in a two-dimensional $M_{\ast}-\mathrm{sSFR}$ plane.

The K-S test is not suitable for comparing the two distributions in this case. Instead, we define a weight function corresponding to the form presented in Equation \ref{eqn:gwrate} to the distribution of control sample. To this end, we normalize both the weighted distribution and the distribution of short and hybrid GRB host galaxies. We then employ a $\chi^2$ minimization technique to minimize a cost function, which is defined as the sum of the squares of the frequency differences in each bin. For this optimization, we utilize the \texttt{minimize} function from the \texttt{scipy.optimize} package.

A significant challenge in determining the optimized weight function on a two-dimensional plane arises from the limited number of short and hybrid GRB host galaxy samples. While the volume-limited COSMOS sample includes  35,122 objects for $z<0.5$, there are only 39 identified short and hybrid GRB host galaxies. Consequently, the control samples exhibit an almost continuous distribution, whereas the host galaxies are sparsely distributed. This disparity complicates the direct comparison of the two distributions, as it is difficult to reproduce a distribution similar to that of short and hybrid GRBs with any assumed weight function. To mitigate this, we have used a sufficiently large bin size. By setting the bin size to 0.5 dex within the ranges of $7<\log(M_*/M_{\odot})<12$, $-12<\log(\mathrm{sSFR}/\mathrm{yr})<-7$, we ensure that most bins containing main sequence galaxies have at least more than one count, demonstrating a nearly continuous distribution (Figure \ref{fig:2dim}).

The use of a large bin size reduces the resolution of the distribution. If one assumes that such a coarse distribution accurately represents the true underlying distribution, $\chi^2$ minimization could yield a weight function that closely aligns the two distributions with minimal errors. However, given the small sample size of the host galaxies, there is no assurance that a smoothed distribution based on this sample accurately reflects the broader population. This necessitates incorporating this uncertainty into our estimation of errors.

A practical method to address this is employing a statistical analysis of result variability via bootstrapping. We conducted 10,000 bootstrapping iterations, each randomly selecting 20-39 host galaxies to derive the best-fit weight function. This sampling method allows for potential repetitions of the same galaxy within individual samples. The median and standard deviations from these iterations provided us with the optimal coefficients and their associated uncertainties, yielding   $(a, b)$=(0.67±0.14,0.14±0.17). The robustness of our findings persisted even when the iteration count varied between 100 and 1000 or when the sample size was consistently set at 39. Consequently, these outcomes align with the previously estimated BNS merger rates, as shown in Figure \ref{fig:grid_search_comp}. Although the orange and red points are marginally outside the 1$\sigma$ error range, the background p-value distribution suggests that they lie on similar trends.

\subsection{BNS Merger Rate Constant} \label{const}

Until now, we compared the short and hybrid GRB host distribution with the volume-limited COSMOS sample and determined the coefficients in the weight function (Equation \ref{eqn:gwrate}). However, without considering the actual occurrence rate of binary neutron star mergers, we cannot determine the normalization constant $C$ value.

There are previous works on the actual rate of BNS mergers. Among them, \citet{2023ApJ...959...13R} corrected for the effect of the sGRB's opening angle to estimate a beaming-corrected event rate of $R_{true}\approx 362-1789 \ \mathrm{Gpc}^{-3} \ \mathrm{yr}^{-1}$. The higher end of this range is based on the naturally narrow jet opening angles observed in sGRBs, while the lower end accounts for the potential of wider opening angles. This estimate falls within the range predicted for the BNS merger rate from the gravitational waves \citep{2020LRR....23....3A}.

In the volume-limited COSMOS sample, if we assume that faint galaxies have a negligible contribution to the BNS merger rate, the sum of the expected log($n_{BNS}$/Gyr) for each galaxy should match the number of BNS mergers expected in the cosmic volume. After rejecting areas contaminated by bright stars, the effective area of the COSMOS field is 1.27 deg$^2$, and the corresponding comoving volume is $8.747\times10^{-4}$ Gpc$^{3}$ ($z<0.5$). Multiplying this by $R_{true}$ gives us the expected BNS mergers in this volume per year.

By incorporating the optimal coefficients obtained in Section~\ref{sec:merger_rate} (as shown in Figure \ref{fig:grid_search}), along with the values of $\log(M_*/M_{\odot})$ and $\log(\mathrm{sSFR}/\mathrm{yr})$ for each galaxy in the volume-limited COSMOS sample, we can define the expected $\log(n_{\mathrm{BNS}}/\mathrm{Gyr})$ as a function of constant $C$:

\begin{equation} \label{eqn:Ceqn}
\sum_{i}^{k}{10^{a\times\log(M_{\ast})_{i} + b\times\log(\mathrm{sSFR})_{i} + C}}=R_{true}\times V_{z<0.5}
\end{equation}

Here, $k$ denotes the total number of galaxies in the volume-limited sample, and $V_{z<0.5}$ represents the comoving volume within $z = 0.5$ over the COSMOS survey field of view. We applied the \texttt{fsolve} function from the \texttt{scipy.optimize} library to solve Equation \ref{eqn:Ceqn}. The result is presented in the \ref{eqn:final}. Given our method of calculating $R_{true}$, the value of $-0.177$ can feasibly be interpreted as reflecting the BNS merger rate.

\begin{equation} \label{eqn:final}
\begin{aligned}
\log(n_{\mathrm{BNS}}/\mathrm{Gyr}) &= 0.86 \times \log(M_{\ast}/M_{\odot}) \\
&\quad + 0.44 \times \log(\mathrm{sSFR/yr}) \\
&\quad + [0.163, 0.857]
\end{aligned}
\end{equation}

To validate this approach to estimate the $C$, we substituted the BNS merger rate coefficients at $z=0.1$ from \citet{2020MNRAS.491.3419A} into the equation, yielding $\log(n_{\mathrm{BNS}}/\mathrm{Gyr}) = 0.800 \times \log(M_{\ast}/M_{\odot}) + 0.323 \times \log(\mathrm{SFR/M_{\odot} yr^{-1}}) + [-3.625, -2.931]$. This range encompasses the constant value of $-3.555$ calculated in the original paper, thereby corroborating the efficacy of our methodology in determining the unit of the BNS merger rate.

\section{Discussion} \label{sec:discussion}

In this section, we will discuss the results we have drawn in the previous section. In the subsection~\ref{sec:bns_discussion}, we will discuss the stellar mass contribution for the merger rate expectation regarding the delay time distribution. In the subsections \ref{sec:kn_discussion} and \ref{sec:sample_discussion}, the behaviors of host galaxies of KNe candidates subgroups and potential biases that could affect the sample and analyses will be discussed. Finally, we will mention other properties of galaxies that could be considered in future works \ref{sec:metal_age_discussion}.

\subsection{Merger Rate Dependence on the Stellar Mass} \label{sec:bns_discussion}

From the mass distribution of short and hybrid GRB host galaxies, we have deduced that the BNS merger rate deviates from a scenario linearly proportional to stellar masses. Instead, our findings suggest that a combined weight function of stellar mass and sSFR more accurately replicates the observed mass distribution of short and hybrid GRB host galaxies.

Figure \ref{fig:grid_search} displays the distribution of K-S test p-values for the comparison between the weighted distribution of galaxies and that of the host galaxies. The optimal coefficients were determined to be  $(a, b) = (0.86^{+0.18}_{-0.18}, 0.44^{+0.34}_{-0.27})$. In contrast, the simulation results from \citet{2020MNRAS.491.3419A} estimated the local universe BNS merger rate (at $z=0.1$) as $\log(n_{\mathrm{BNS}}/\mathrm{Gyr}) = 0.800 \times \log(M_*/M_{\odot}) + 0.323 \times \log(\mathrm{SFR/M_{\odot} yr^{-1}}) - 3.555$. This equates to $(a, b) = (1.123, 0.323)$, significantly diverging from the error range of our results. This discrepancy highlights the pivotal role of star-forming galaxies, suggesting that stellar mass alone may be less influential in the context of predicting the BNS merger rate.

 The significant contribution of star-forming galaxies to BNS merger host galaxies aligns with more recent simulation studies. \citet{2022MNRAS.514.2716M} predicted, using synthetic binary models and hydrodynamic simulations, that local BNS merger host galaxies will exhibit blue color and high star-forming activity. Similarly, \citet{2022MNRAS.509.1557C} emphasized the importance of stellar mass, SFR, and metallicity of host galaxies in determining the BNS merger rate, predicting that local host galaxies will be dominated by late-type galaxies. Studies such as \citet{2020MNRAS.499.5220M} and \citet{2020ApJ...905...21A}, which predict the characteristics of hosts based on various delay time models, also suggest that local host galaxies could be dominated by star-forming galaxies with BNS mergers with short delay times. However, for nearby host galaxies like NGC 4993, it is necessary to consider scenarios where the effects of galaxy evolution and progenitors with long delay times result in mergers occurring in massive galaxies.

\begin{figure}[b]
\begin{center}
\includegraphics[width=\columnwidth]{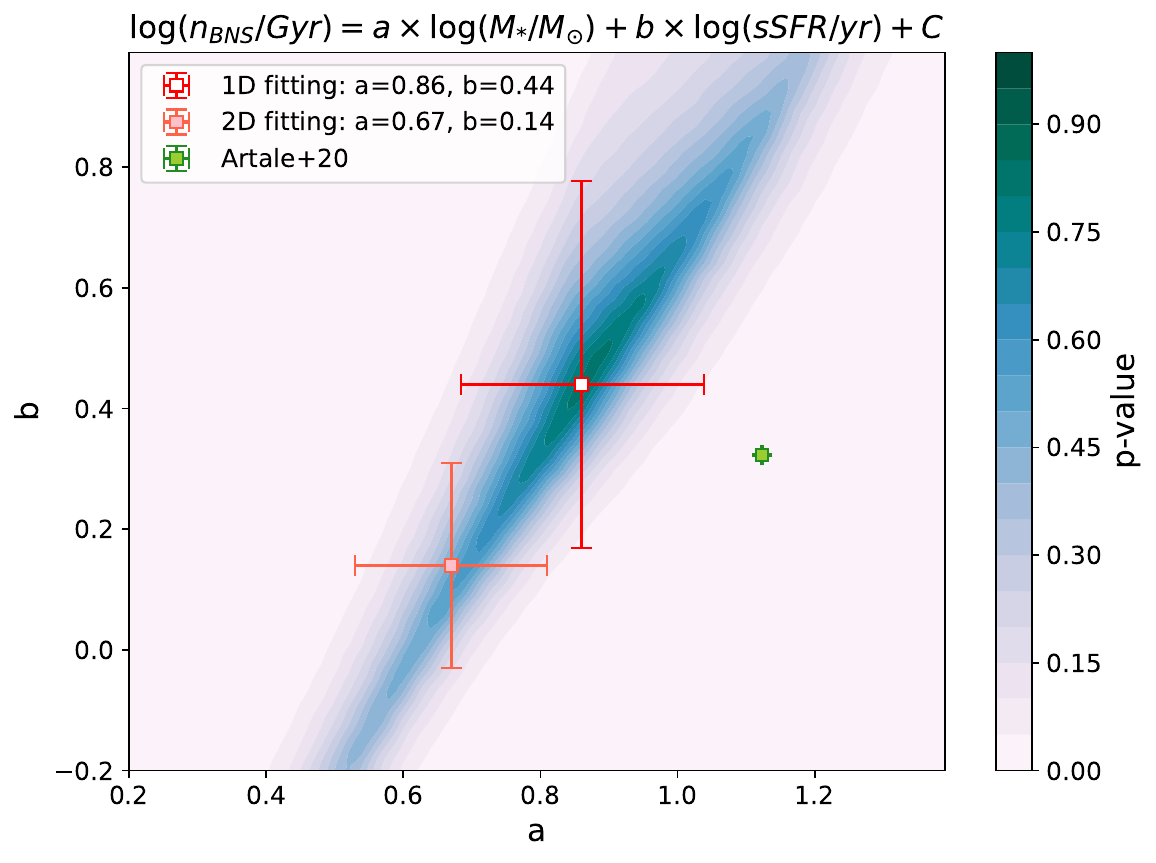}
\caption{The comparisons of the BNS merger rate coefficients by overlaying the outcomes of Figure \ref{fig:2dim} and \citet{2020MNRAS.491.3419A} on the merger rate coefficient parameter space. Here, the optimized parameters $a$ and $b$ are defined in the relationship of  $\log(n_{\mathrm{BNS}}/\mathrm{Gyr}) = a \times \log(M_{\ast}/M_{\odot}) + b \times \log(\mathrm{sSFR/yr}) + C$. 
\label{fig:grid_search_comp}}
\end{center}  
\end{figure}

If we disregard the sSFR component and fix the $b$ coefficient to 0, the best-fit value of $a$ becomes  $0.62\pm0.06$. This implies assuming $\log(n_{\mathrm{BNS}}/\mathrm{Gyr}) = 0.62 \times \log(M_{\ast}/M_{\odot}) + C$, and aligns with the conclusion that the significance of stellar mass should be considered to a lesser extent than 1. Compared to utilizing the merger rate function that incorporates both stellar mass and sSFR, considering only the stellar mass component diminishes the similarity to the short and hybrid GRB host galaxy distribution. Correspondingly, the p-value of the K-S test declines from 0.82 to 0.49, thereby reinforcing the necessity of including the sSFR in the second term of the analysis.

As the tendency between $a$ and $b$ depicted in Figure~\ref{fig:grid_search_comp} implies, the more contribution we attribute to stellar mass, the more significant sSFR component should be in predicting the BNS merger rate. Therefore, compared to a merger rate prediction dominated by mass, incorporating sSFR regulates larger population of passive and massive galaxies, which typically exhibit negligible sSFR.

In actual follow-up observation scenarios, assigning linear weight to stellar mass could reduce observational efficiency. For example, if a $10^{11} M_{\odot}$ galaxy is located in a peripheral region with a lower volume probability, it might be assigned a higher priority because it can easily outweigh the combined mass of galaxies in other pointings. In contrast, using $M_{\ast}^{0.62}$ as a weight would likely yield results that more appropriately reflect the overall distribution of galaxies.

\subsection{Host Galaxies of the Kilonova Candidates} \label{sec:kn_discussion}

One compelling piece of evidence in probing BNS mergers involves examining sGRBs that display a kilonova (KN) component. Their late-time infrared emission excess has been suggested as a sign of KNe. This phenomenon is attributed to the decay of opaque heavier elements that persist until late times \citep{2019ApJ...876..128K}.

As detailed in Section~\ref{sec:kn_sample}, several GRBs have been observed to exhibit KN emissions in their afterglows. Referring to previous studies, we have identified the host galaxies of 12 KN candidates at $z<0.5$. Analyzing the properties of these host galaxies enables a more direct assessment of the BNS merger environment.

\begin{figure*}
\begin{center}
\includegraphics[width=\linewidth]{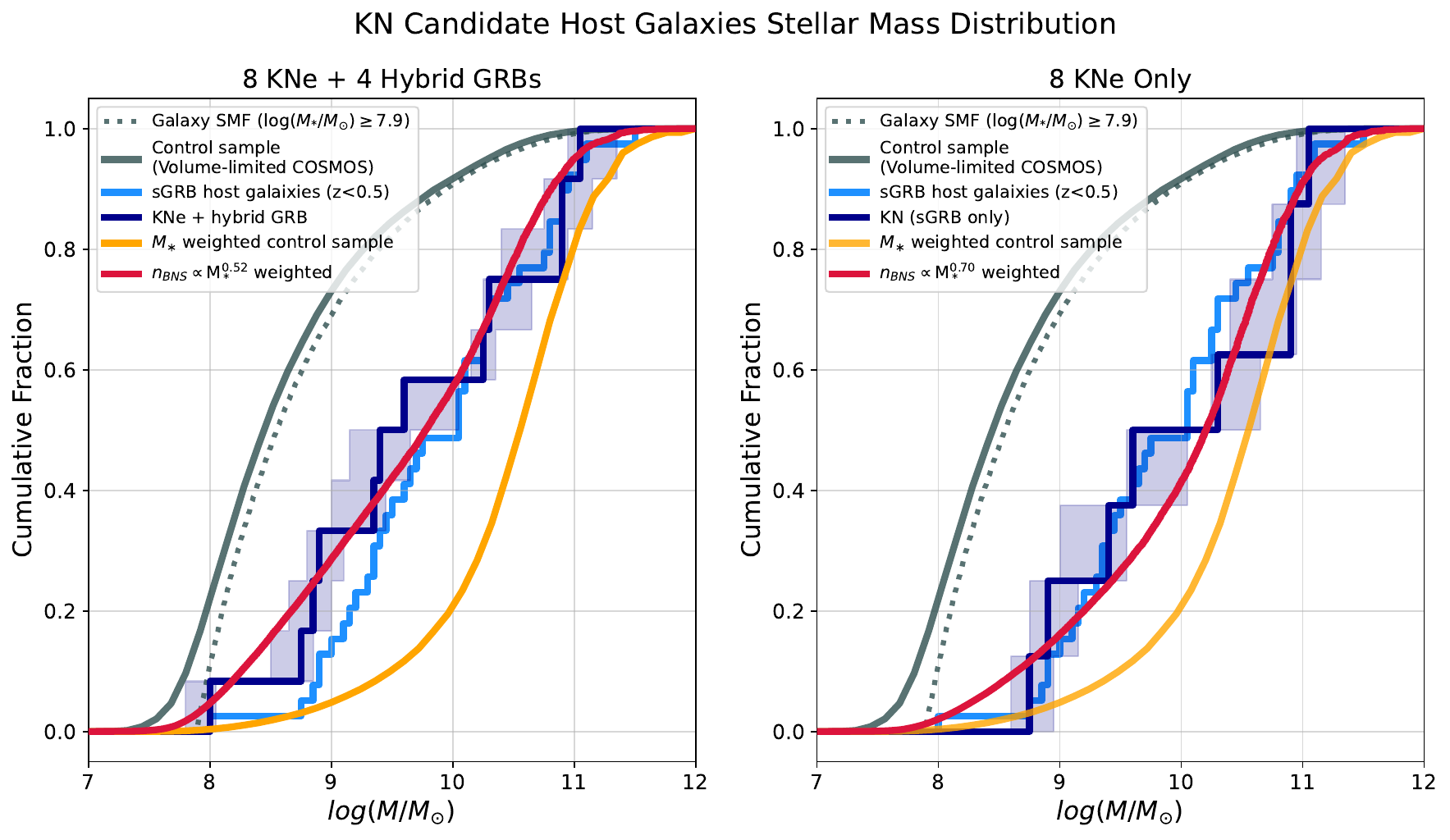}
\caption{(\textbf{Left}) The stellar mass cumulative distribution, similar to Figure \ref{fig:cummulative} but specifically limited to kilonova candidates host galaxies. The navy line represents the mass distribution of 12 kilonova host galaxies, and the shaded region indicates the uncertainty range in stellar mass estimates. The red line indicates the distribution of control sample weighted by  $\log(n_{\mathrm{BNS}}/\mathrm{Gyr}) = 0.52 \times \log(M_{\ast}/M_{\odot}) + C$, optimally reproducing the observed kilonova host distribution. (\textbf{Right}) The solid navy line illustrates the mass distribution when excluding four additional events associated with long GRBs, and the corresponding optimally reproduced weighted distribution is denoted by the red line. For this case, the optimal weight function is determined as  $\log(n_{\mathrm{BNS}}/\mathrm{Gyr}) = 0.70 \times \log(M_{\ast}/M_{\odot}) + C$, suggesting more massive distribution compared to the rest of sGRB host galaxies.
\label{fig:kn_cummulative}}
\end{center}
\end{figure*}

Figure \ref{fig:kn_cummulative} presents the stellar mass distribution of the host galaxies of KN candidates. The distribution for the 12 KN candidates does not show a remarkable difference from that of other sGRB host galaxies. This similarity may suggest that short and hybrid GRB host galaxies are a representative sample of the BNS merger environment. The KN host galaxy's mass distribution also underscores the importance of not overlooking low mass galaxies when studying BNS merger host galaxies.

However, it is important to note that the KN candidates mentioned here include not only typical sGRB afterglows but also hybrid GRBs (kilonova with long GRBs) like GRB 060614, GRB 060505, GRB 211211A, and GRB 211227A. These instances are hypothesized to be binary systems with neutron stars, but there is a possibility that they represent a unique group, such as binary mergers involving neutron stars and white dwarfs \citep{2023ApJ...947L..21Z}.

When we exclude the hybrid GRBs and analyze the stellar mass distributions of the nine host galaxies where KNe were securely associated with sGRBs, the host galaxies appear marginally more massive, as shown in Figure \ref{fig:kn_cummulative}.  To address this effect, we assumed that the KN rate can be estimated as a function of stellar mass. We found that the optimal weight function is $M_{\ast}^{0.52}$ for KN candidates and hybrid GRB host galaxies, and $M_{\ast}^{0.70}$ when limited to the 8 KN candidate host galaxies. However, this difference is not statistically significant enough to support the null hypothesis that hybrid GRBs and the rest of KNe are distinct populations, as evidenced by the K-S test result ($P_{KS}=0.25$).

\subsection{Potential Biases from Samples and Analyses} \label{sec:sample_discussion}

We explored the BNS merger environment by comparing 39 short and hybrid GRB host galaxies at $z<0.5$ with a volume-limited sample from the COSMOS field galaxies. However, among the 39 host galaxies used, some have uncertain associations, raising the need to assess their impact on our results. Most samples with $P_{CC}>0.10$ were excluded by selecting local samples, but four cases without afterglow detection (GRB 050906, GRB 070810B, GRB 080121, GRB 100216A) remain highly uncertain. These occurred near bright galaxies, and the bright galaxies were chosen as host candidates based on the low probability of random spatial overlap with the burst positions \citep{2020MNRAS.492.5011D}. If these cases are biased towards a certain part of the massive distribution, it might influence in our analysis.

\begin{figure}
\begin{center}
\includegraphics[width=\columnwidth]{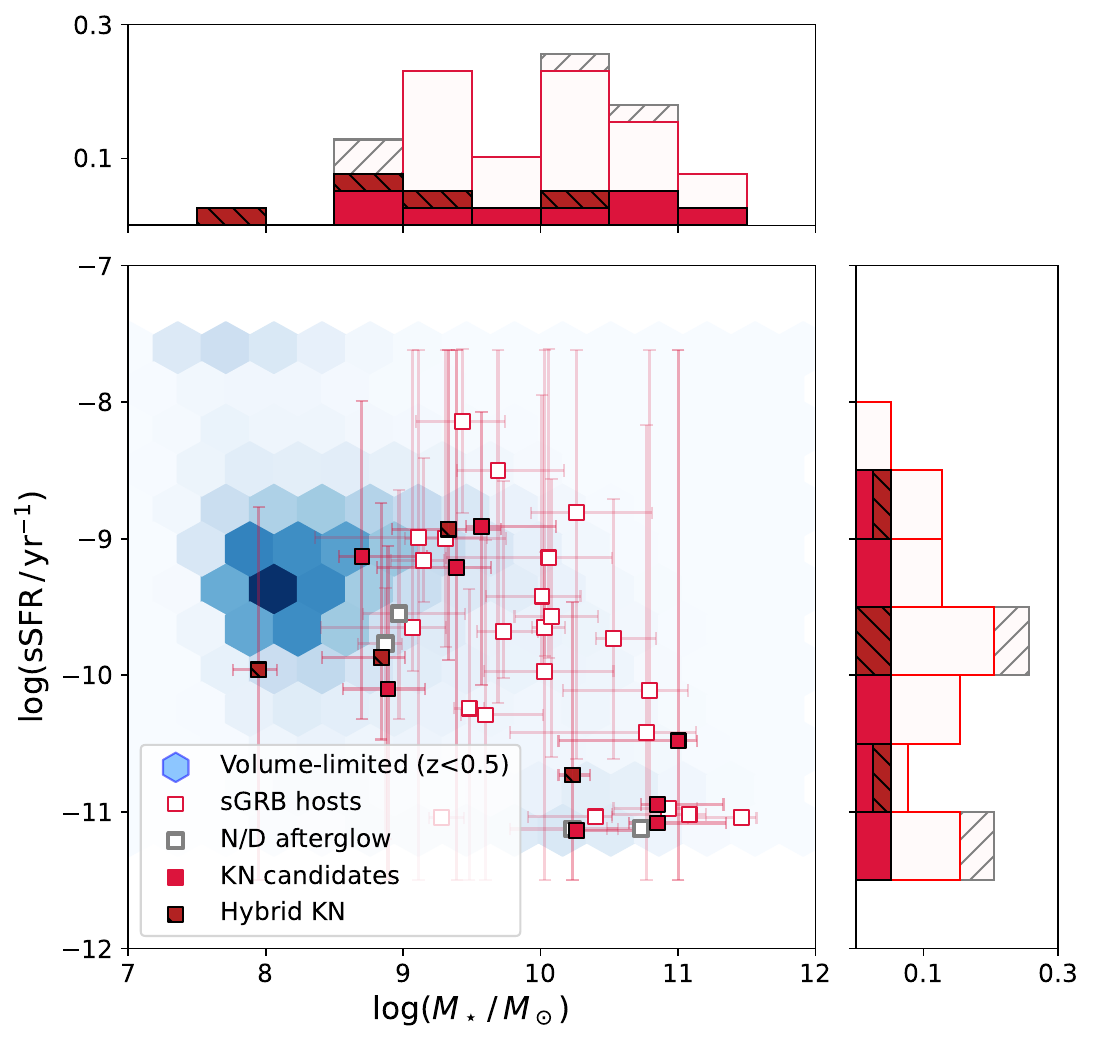}
\caption{The distribution of 39 short and hybrid GRB host galaxies in the $\log(M_{\ast}/M_{\odot})-\log(\mathrm{sSFR/yr})$ plane. The blue hexagon bin in the background represents the distribution of volume-limited COSMOS sample with $z<0.5$. Red-filled points denote instances where kilonova emission features were observed, with hybrid GRBs that occurred alongside LGRBs distinctly marked with hatching. The samples outlined in gray are 4 samples suggested by \citet{2020MNRAS.492.5011D}, without detection of afterglow emissions, thus marking them as highly uncertain. The histograms along each axis represent the distributions of stellar mass and sSFR for the sample without afterglow detections or the one that is only specifically limited to cases showing kilonova features. Each of the distributions shows no statistically significant difference from the overall sample.
\label{fig:M-sSFR distribution}}
\end{center}
\end{figure}

Nonetheless, as illustrated in the upper panel of Figure \ref{fig:M-sSFR distribution}, the stellar masses of host galaxies without afterglow detection are not biased towards a particular mass range. Upon excluding the afterglow non-detection cases and reapplying our grid searching method, the results  (a, b)=(0.82, 0.39) remained similar. This indicates that the potential sample bias due to uncertain host associations is minimal for $z<0.5$ short and hybrid GRB host galaxies.

A challenging aspect of sample bias in the local universe relates to hostless galaxies. Simulations by \citet{2022MNRAS.514.2716M} indicate that progenitors from low mass galaxies, due to weaker gravitational potentials, may more readily escape and produce fainter afterglows. This makes such galaxies less likely to be observed as sGRB host galaxies. Predictions suggest that hostless events might comprise approximately 16-35\% of BNS mergers.

Recently, \citet{nugent2023population} posits that a significant portion of high-redshift sGRB host galaxies belong to a dwarf host population. This finding may imply that our observed host properties could be biased towards brighter galaxies, differing from the actual BNS merger environments. However, to fully trust this argument, one must also consider that sGRB events connected to faint galaxies are very rare in the local universe. In our sample, the only confirmed faint host galaxy is that of GRB 060614, identified as a hybrid LGRB event. Given this, we conclude that current evidence is insufficient to ascertain the contribution of hostless or faint galaxies to the true BNS merger environment.

On the other hand, the range of parameters we used in the equation may have caused differences in the form of the equation. We set a lower limit for sSFR estimated from SED fitting because it is not physically meaningful to analyze the SFR of quiescent galaxies with values close to 0. Among various criteria for classifying quiescent galaxies, we used Equation \ref{eqn:passive}. This is similar to setting a lower limit of sSFR to about $10^{-11} \mathrm{yr}^{-1}$, and it is also similar to the $NUVrJ$ color-color selection used by \citet{2023A.A...677A.184W}.

However, the definition of a passive galaxy can sometimes vary, being set as $\mathrm{sSFR} > 1/[3 \times t(z)]$ or $ > 1/t(z)$, similar to establishing a lower sSFR limit of $10^{-10.5}\ \mathrm{yr}^{-1}$ or $10^{-10}\ \mathrm{yr}^{-1}$, respectively. When the sSFR lower limit is defined as $1/[3 \times t(z)]$, the best-fit coefficients are calculated to be $(a, b) = (0.92^{+0.09}_{-0.24}, 0.62^{+0.19}_{-0.44})$. This shift suggests that as the sSFR cut for quiescent galaxies is assumed to be higher, the value of the coefficient $b$ also shifts upwards. Consequently, this highlights the importance of retaining the definition of passive galaxies and their parameters to appropriately apply the resulting equation.

Finally, the definition of the control sample can also influence the results. To ensure completeness down to $\log(M_{*}/M_{\odot}) = 7.9$, we set the absolute magnitude cut at $M_r < -16.22 \ \mathrm{mag}$. If we lower this threshold to $M_r < -14 \ \mathrm{mag}$ to include fainter galaxies, the control sample would expand from  35,122 to 77,434 galaxies. In this scenario, the best-fit coefficients change to approximately $(a, b) = (0.95^{+0.15}_{-0.16}, 0.56^{+0.33}_{-0.24})$. This change reflects the inclusion of more galaxies with lower mass and sSFR values, thus assigning a higher weight to them. However, the ratio between $a$ and $b$ remains nearly constant while varying $M_r$ cut, suggesting that from a prioritization perspective, the outcome is maintained.

\subsection{Other Host Galaxy Parameters} \label{sec:metal_age_discussion}

Our analysis primarily focuses on host galaxies' stellar mass and sSFR to inspect the BNS merger environment. However, we should acknowledge that other properties could also affect the predicted merger rate. 

One such factor is the age of the stellar population. Galaxies harboring older stars are anticipated to exhibit a reduced merger rate, a trend influenced by the delay time distribution (DTD) effect. However, as we presented in Figure \ref{fig:refcomp}, stellar population age is not a well-constrained parameter.

Simulation studies, such as those conducted by \citet{2018MNRAS.481.5324M}, \citet{2020MNRAS.491.3419A} and \citet{2022MNRAS.509.1557C}, have underscored the role of host metallicity in progenitor formation, indicating a likely positive correlation with the BNS merger rate. This correlation arises because the metallicity influences the initial mass of the compact binary progenitor systems. 

Estimating galaxy parameters through the SED fitting is challenging with the currently limited broad-band photometry information for some of our datasets and the SED fitting model. 
The metallicity resolution is constrained to just four distinct levels in the stellar population library by \citet{2003MNRAS.344.1000B} employed for the SED fitting. Furthermore, there is a degeneracy among the stellar age, the metallicity, and the reddening. For example, the impact of aging is approximately 1.5 times the effect induced by metal abundance \citep{1994ApJS...95..107W}.

Future investigations, with additional photometry datasets and spectroscopic data containing emission and absorption line measurements, should facilitate the estimation of galaxy parameters.

\subsection{Application of Our Results to Sutdies of GW Events and Other Transients}

Our findings offer practical guidelines for follow-up observations of gravitational wave events. Our result suggests that GW event follow-up observation can be efficiently prioritized if both mass and star formation are considered.  Assuming the BNS merger rate follows $M_{\ast}^{0.62}$ instead of being directly proportional to stellar mass, the effects are as follows. We would need to observe a larger number of galaxies to cover the same amount of probability, and we would be more strongly affected by bias due to the incompleteness of the reference catalog. Nevertheless, we need to take into account the observational evidence that significant fraction of sGRBs and KN-like emissions is found in star-forming galaxies. Our results provide quantitative guidelines for tiling or targeting observations for gravitational wave follow-up, and are practical for developing observation strategies and systematically utilizing facilities. For example, our BNS merger rate prescription can be applied each galaxy in the GW localization area. Then, the galaxies can be sorted in order of decreasing merger rate for more efficient galaxy-target observation (e.g., \citet{2024ApJ...960..113P}). Future observations of the EM-producing GW events should tell whether our BNS merger rate estimator is superior to BNS merger rate estimators based on stellar mass only.

The methodology employed in this study to assess host galaxy candidates within the GW probable volume can be applied retroactively. For instance, by weighting the probabilities of galaxies within the localization volumes using our BNS merger rate prescription, we can estimate host galaxy distance distributions for these gravitational wave events. Accumulated data from such cases could lead to statistical analyses of cosmological parameters in future research.

This approach is also applicable to predicting other types of transients. Studies that estimate transient rates by measuring how host galaxy properties deviate from those of general galaxies have been actively conducted for supernovae. Recently, \citet{liang2024luminosity} analyzed the stellar mass function of host galaxies and found that for core-collapse supernovae, weighting by SFR is crucial, while for type Ia supernovae, both SFR and stellar mass need to be considered. When studying recently discovered transients (e.g., GRBs, tidal disruption events and fast radio bursts), our approach of comparing galaxy properties within volume-limited samples will be effective for probing their uncertain progenitor systems and defining the characteristics of their host galaxies.

\section{Conclusion} \label{sec:conclusion}

In this study, we studied the galactic environment where BNS mergers occur, focusing on the host galaxies of short and hybrid GRBs. We compiled a dataset of the GRB host galaxies, gathering multi-wavelength broad-band photometry and spectroscopic redshift data from previous studies and archival images. This data set was utilized for the SED fitting to deduce the properties of the host galaxies.

Our analysis was confined to 39 host galaxies at $z<0.5$. We compared these with a volume-limited sample from galaxies in the COSMOS field, achieving a complete dataset down to $10^{7.9} M_{\odot}$ at $z<0.5$. Figure \ref{fig:cummulative} reveals that the stellar mass distribution of short and hybrid GRB host galaxies deviates from the volume-limited COSMOS sample, where the number count is weighted by each galaxy's stellar mass. This would not be true if the BNS merger rate were proportional to the host galaxy stellar mass. Instead, we found that the BNS merger rates, which successfully reproduce the observed distribution of host galaxies, can be expressed by the following equation: 
$\log(n_{\mathrm{BNS}}/\mathrm{Gyr}) = 0.86 \times \log(M_{\ast}/M_{\odot}) + 0.44 \times \log(\mathrm{sSFR/yr}) + 0.857$.

Dividing the short and hybrid GRB sample into several subsamples does not affect the conclusion. The stellar mass distributions of sGRBs with ambiguous host associations are consistent with those of sGRBs with clearer host identifications. Additionally, events exhibiting kilonova signatures show host characteristics similar to those of other sGRB host galaxies. Excluding hybrid GRB events from the kilonova signature sample resulted in a marginally more massive host distribution, but not enough to draw definitive conclusions.

 With the expected improvements in GW detection technologies, more GW events will be discovered in future that would be accompanied by KNe. Our results on the BNS merger rate estimator can serve as an important element for designing efficient search methods to find optical counterparts of  GW events from BNS mergers and eventually help promote multi-messenger astronomy involving GW. Our methodology of estimating the merger rate can be applied to host galaxies of other types of transients such as FRBs to understand the environment of such transients.

\begin{acknowledgments}
We express our gratitude to the anonymous referee for their valuable comments and suggestions that significantly improved the manuscript. We also thank Donggeun Tak for proofreading our paper and providing insightful feedback. This work was supported by the National Research Foundation of Korea (NRF) grants, No. 2020R1A2C3011091 and No. 2021M3F7A1084525, funded by the Korean government (MSIT).
\end{acknowledgments}

\appendix

\renewcommand{\thefigure}{A\arabic{figure}}
\renewcommand{\thetable}{A\arabic{table}}
\setcounter{figure}{0}
\setcounter{table}{0}

\hspace*{-1cm}
\scriptsize 
\begin{longtable}{p{1.9cm}|>{\centering\arraybackslash}p{1.7cm}>{\centering\arraybackslash}p{1.7cm}|>{\centering\arraybackslash}p{1.4cm}|>{\centering\arraybackslash}p{2.5cm}>{\centering\arraybackslash}p{0.9cm}>{\centering\arraybackslash}p{1.7cm}|p{4cm}}
\caption{Photometric data for short and hybrid GRB host galaxies}\label{tab:hostphot} \\

\toprule
    Target &            RA &           Dec &                redshift &        Instrument & Filter &          ABmag &                      Reference$^{a}$ \\
\midrule
\endfirsthead

\toprule
    Target &            RA &           Dec &                redshift &        Instrument & Filter &          ABmag &                      Reference$^{a}$ \\
\midrule
\endhead
\midrule
\multicolumn{8}{r}{{Continued on next page}} \\
\midrule
\endfoot
\bottomrule
\endlastfoot

GRB050509B & 12:36:12.87 & 28:58:58.84 &                 $0.225$ &              SDSS &      u & 20.32$\pm$0.13 &    \citet{2010ApJ...725.1202L} \\
           &             &              &                         &               PS1 &      g & 18.41$\pm$0.07 &    \citet{2016arXiv161205560C}$^{*}$ \\
           &             &              &                         &               PS1 &      r & 17.30$\pm$0.07 &    \citet{2016arXiv161205560C}$^{*}$ \\
           &             &              &                         &               PS1 &      i & 16.80$\pm$0.08 &    \citet{2016arXiv161205560C}$^{*}$ \\
           &             &              &                         &               PS1 &      z & 16.53$\pm$0.06 &    \citet{2016arXiv161205560C}$^{*}$ \\
           &             &              &                         &               PS1 &      y & 16.41$\pm$0.16 &    \citet{2016arXiv161205560C}$^{*}$ \\
           &             &              &                         &             2MASS &      J & 16.16$\pm$0.14 &    \citet{2010ApJ...725.1202L} \\
           &             &              &                         &             2MASS &      H & 15.84$\pm$0.18 &    \citet{2010ApJ...725.1202L} \\
\end{longtable}

\newcounter{customfig}
\setcounter{customfig}{\value{figure}} 

\begin{figure*}[p]
\centering
\includegraphics[width=\textwidth]{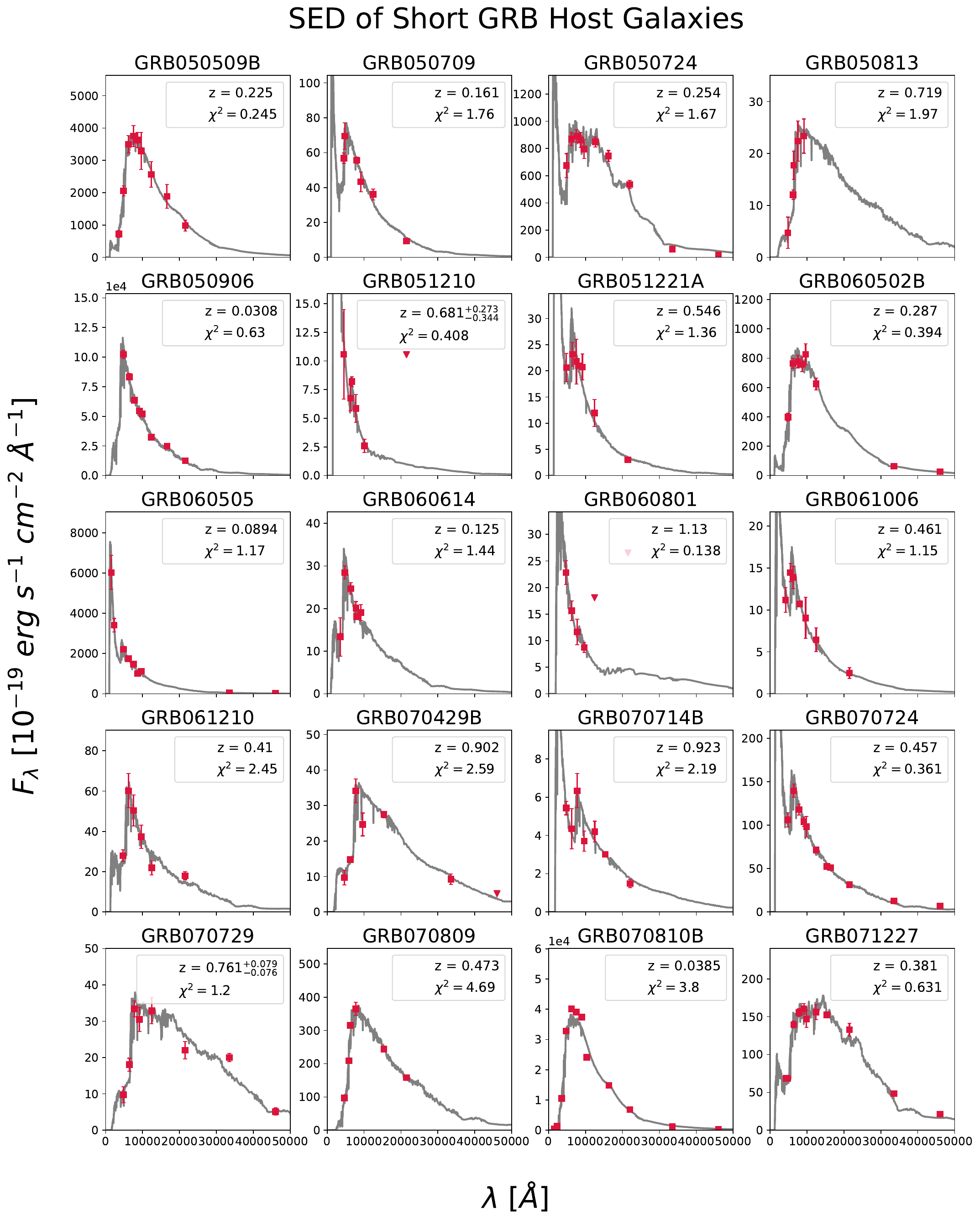}
\caption{Best fit SED models and photometric data points of short and hybrid GRB host galaxies. (Part 1)}
\label{fig:sed}
\end{figure*}

\addtocounter{customfig}{1} 
\setcounter{figure}{\value{customfig}} 

\begin{figure*}[p]
\centering
\includegraphics[width=\textwidth]{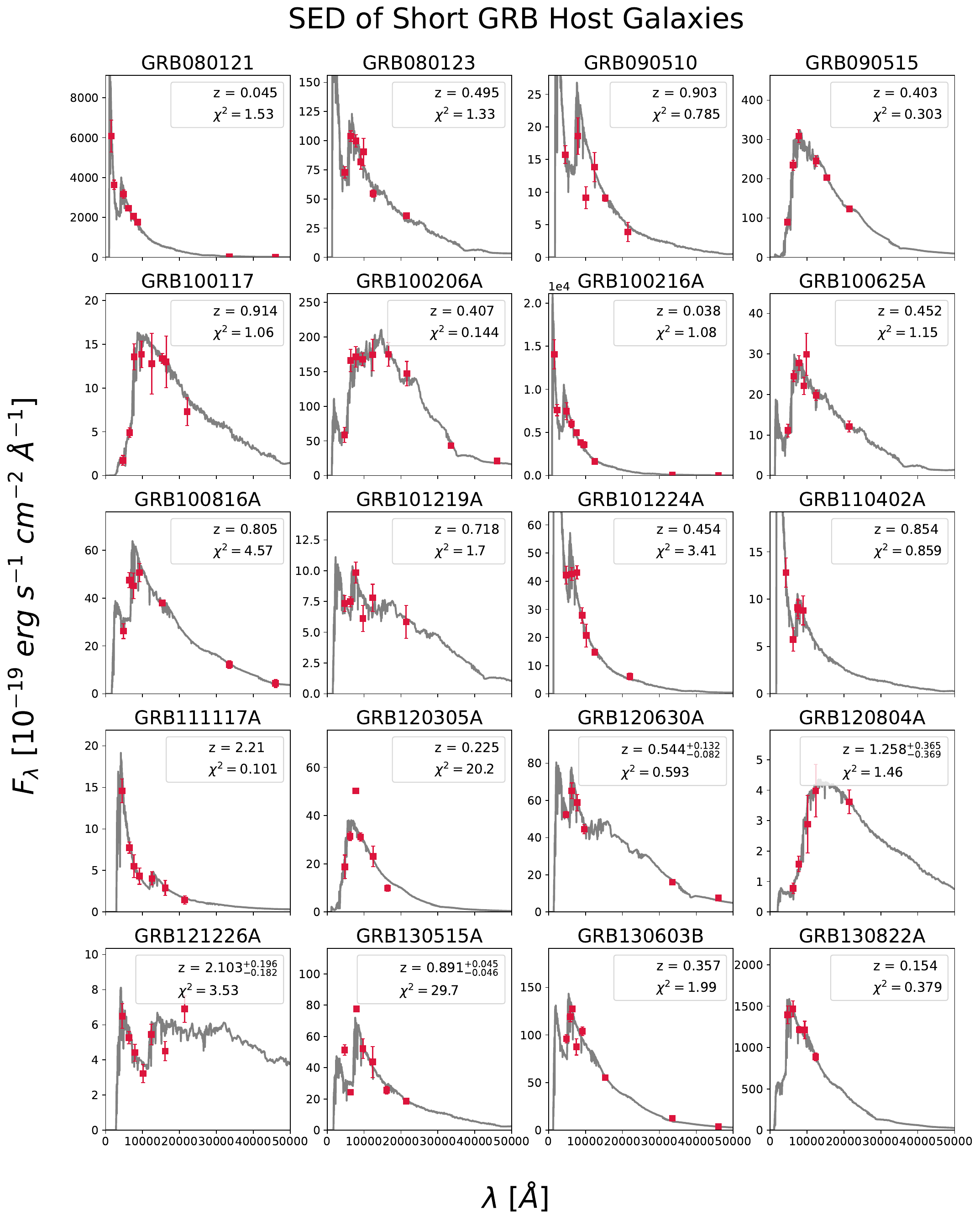}
\caption{Best fit SED models and photometric data points of short and hybrid GRB host galaxies. (Part 2)}
\label{fig:sed2}
\end{figure*}

\addtocounter{customfig}{1} 
\setcounter{figure}{\value{customfig}} 

\begin{figure*}[p]
\centering
\includegraphics[width=\textwidth]{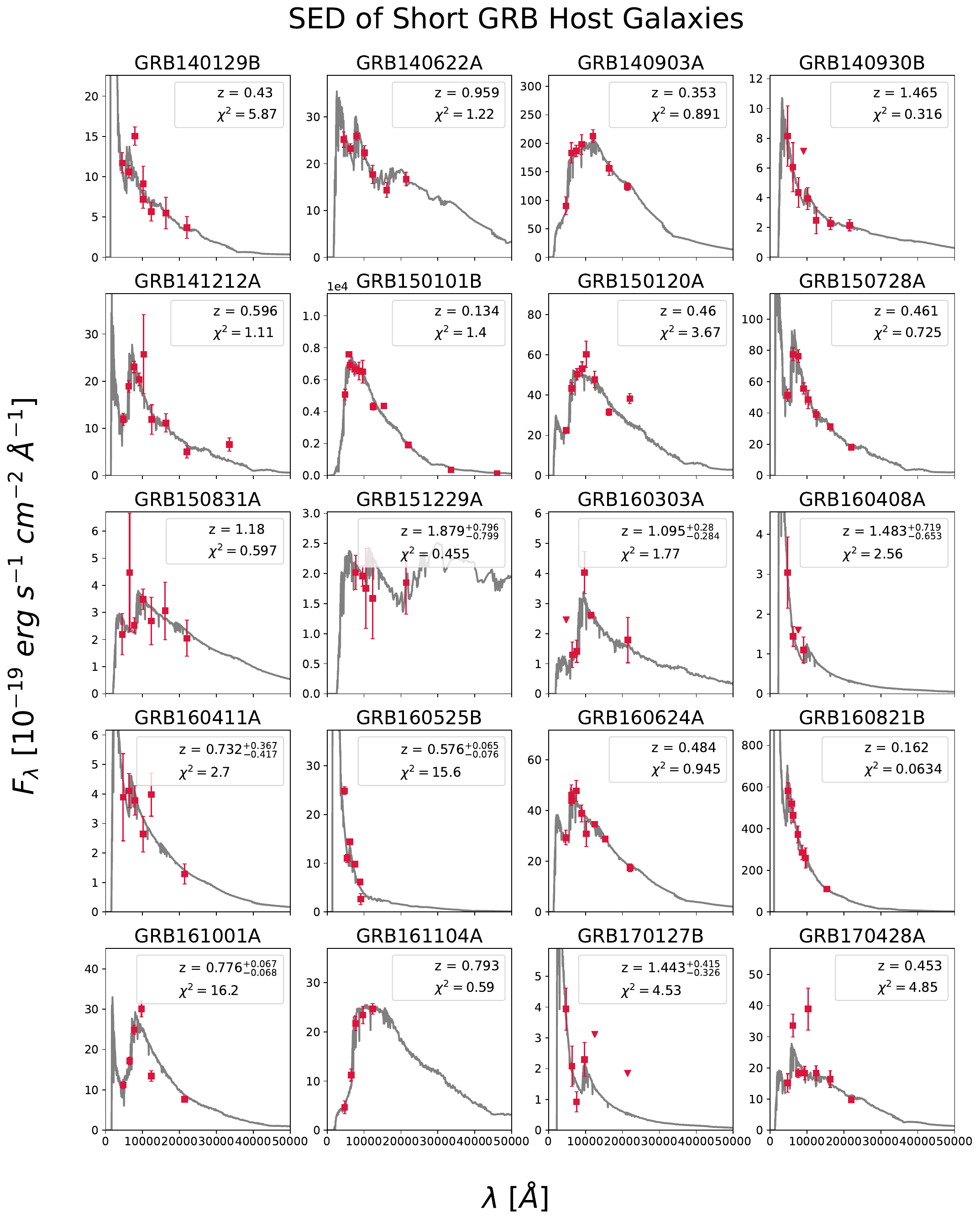}
\caption{Best fit SED models and photometric data points of short and hybrid GRB host galaxies. (Part 3)}
\label{fig:sed3}
\end{figure*}

\addtocounter{customfig}{1} 
\setcounter{figure}{\value{customfig}} 

\begin{figure*}[p]
\centering
\includegraphics[width=\textwidth]{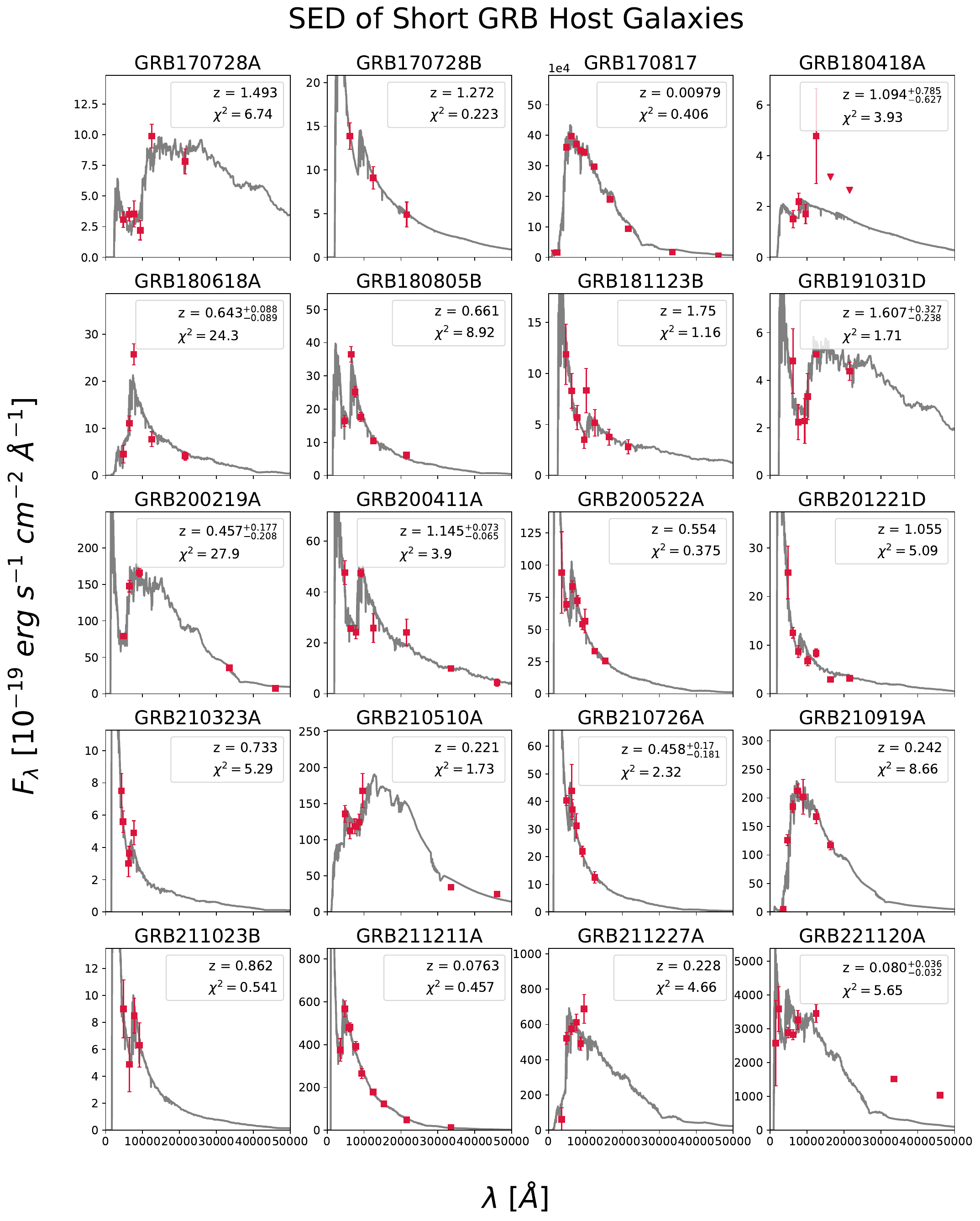}
\caption{Best fit SED models and photometric data points of short and hybrid GRB host galaxies. (Part 4)}
\label{fig:sed4}
\end{figure*}

\setcounter{figure}{\value{customfig}} 

\newpage
\bibliography{hogwarts}{}
\bibliographystyle{aasjournal}

\end{document}